\documentclass[10pt,twocolumn,twoside,,letterpaper,final]{IEEEtran}

\usepackage{subfigmat}

\usepackage{cite}
\ifCLASSINFOpdf
 \usepackage[pdftex]{graphicx}
\else
 \usepackage{epsfig}
\fi
\usepackage[cmex10]{amsmath}
\usepackage{amsfonts}
\usepackage{algorithmic}
\usepackage{algorithm}
\usepackage{array}
\usepackage{mdwmath}
\usepackage[caption=false,font=footnotesize]{subfig}
\usepackage{url}
\ifCLASSOPTIONdraftcls

\else

\fi

\usepackage{bm}

\usepackage{color}

\usepackage{footnote}

\makesavenoteenv{tabular}
\makesavenoteenv{table}

\begin{document}
\title{P-BOOST: Parallel Boosting of Optimal Narrow-Band Direction of Arrival Estimators}

\author{Elio~D.~Di~Claudio, 
 Raffaele~Parisi,~\IEEEmembership{Senior~Member,~IEEE,}
 and~Giovanni~Jacovitti% <-this % stops a space
\thanks{Manuscript received \today; revised Month, Year.}
\thanks{Elio D. Di Claudio and Raffaele Parisi are with the Department of Information Engineering, Electronics and Telecommunications, University of Rome ``La Sapienza'', via Eudossiana 18, 00184 Rome, Italy. (e-mail: \{elio.diclaudio, raffaele.parisi\}@uniroma1.it).}
\thanks{G. Jacovitti, retired, was with the Department of Information Engineering, Electronics and Telecommunications, Sapienza University of Rome, Rome 00184, Italy (e-mail: gjacov@infocom.uniroma1.it).}}
%\thanks{Digital Object Identifier 10.1109/TASL.201X.1234567}
%

% ^------------^------------^----Do not want these spaces!

\markboth{Proposal2019}%
{Di Claudio \MakeLowercase{\textit{et al.}}: P-BOOST: Parallel boosting of optimal narrow-band direction of arrival estimators}
% *** name in the headers of peer review papers. ***



\maketitle

\begin{abstract}
Optimal Maximum Likelihood (ML), narrow-band direction finding cannot be easily initialized in coherent and low signal to noise ratio environments. Sparse under-determined solvers are considered as viable solutions to this problem, since they drastically reduce the dimensionality of the search space by exploiting the array model sparseness. However, because of quantized locations, conventional sparse solvers present some ambiguity problems. In this work, we propose a novel boosting scheme for ML-type estimators, referred to as Parallel BOOSTer (P-BOOST), where a set of generalized MUSIC solutions provides pre-estimates of the directions and the number of coherent paths for arbitrary sensor array geometry and noise covariance. P-BOOST delivers improved and reliable coarse parameter estimates to a further ML or sparse optimization stage even in coherent and/or high noise scenarios. Moreover, its dataflow is highly parallel, which is essential in foreseen remote sensing and telecommunication applications and fully justifies its acronym.
\end{abstract}


\ifCLASSOPTIONpeerreview
 \begin{center} \bfseries EDICS Category: SAM-DOAE, RAS-LCLZ. \end{center}
\fi
\IEEEpeerreviewmaketitle

\section{Introduction}
\label{section:intro}
\IEEEPARstart{T}{he} theory of parametric direction finding was developed along the three last decades \cite{VANTREES02}. Several estimators approach the Cramer-Rao bound (CRB) for narrow-band direction of arrival (DOA) estimation with Gaussian signals and noise. They are based on Maximum Likelihood (ML) \cite{BOHME85} or Weigthed Subspace Fitting (WSF) \cite{STOICA90a,VIBERG91} criteria, that fit respectively the spatial covariance or a Minimum Mean Square Error (MMSE) signal subspace estimate \cite{DEMOOR93}, by searching for the proper combination of steering vectors (vector array frequency responses \cite{SCHMIDT86}) and source parameters. 

However, with the notable exception of MODE \cite{STOICA90a}, that implements WSF by an iterative spatial smoothing \cite{WAX94} and polynomial rooting, valid for uniform linear arrays (ULA) or rectifiable arrays \cite{FRIEDLANDER93,WEISS95}, optimal DOA estimators require a coarse initialization of the source location parameters. In fact their \emph{trust region} \cite{LUENBERGER89} extends only for a fraction of beam-width around the true DOAs \cite{CADZOW90} and they must know the number of paths in the coherent case, not available by spatial covariance rank estimation only \cite{WAX85,NADAKUDITI08}. 

Conventional narrow-band beamforming \cite{VANTREES02} is robust in low signal to noise ratio (SNR) and coherent environments, but sidelobes prevent effective multiple source localization. Minimum Variance (MV) beamforming has better resolution, but it is limited in coherent, narrow-band scenarios \cite{LI05}. Suboptimal DOA estimators, such as MUSIC \cite{SCHMIDT86} or ESPRIT \cite{ROY89}, break down for coherent sources.

Therefore, in the last decade there has been an increasing interest for \emph{sparse linear under-determined solvers}, that mimic integral field equations \cite{MASSA15} and search for \emph{localized energy solutions} \cite{GORODNITSKY97} over a spatially sampled manifold of steering vectors, herein referred to as \emph{codebook}. A sparse solution involving only few codebook elements is matched in the ML or WSF sense to array output measurements \cite{MALIOUTOV05}. 

The first attempt is perhaps due to Cadzow \cite{CADZOW90}, that searched the sequence of steering vectors minimizing the WSF criterion at each step. This fast \emph{sequential beamforming}, referred to as \emph{Alternate Projection}, is today recognized as a refined \emph{Orthogonal Matching Pursuit} (OMP) \cite{MALLAT93} and initializes a local descent in the DOA parameter space. 

Since the signal subspace OMP \cite{CADZOW90} relies on conventional beamforming, it is highly biased by sidelobes for multiple and coherent sources. As a consequence, OMP exhibits a strong excess WSF error at each step and there is not any statistically sound way of stopping iterations at the correct number of sources. This task is left to cumbersome combinatorial tests after local WSF optimization for each hypothesis \cite{VIBERG91,GERSHMAN99}.

Other \emph{sparse} solvers useful for coarse DOA initialization are the FOCal Underdetermined System Solution (FOCUSS) \cite{GORODNITSKY97} and the $L_1$ penalized solution \cite{MALIOUTOV05}, that were applied in several environments related to direction finding, such as magneto-encephalography (MEG). These methods penalize non-sparse solutions at each iteration and were later interpreted and enhanced as \emph{sparse Bayesian solvers} (SBSs), where the prior is posed on source amplitudes \cite{WIPF07}.

In this paper, it is stressed that classical DOA parametric estimators realize already a sparse and optimal representation of the array signal model in the \emph{continuous} DOA domain \cite{FANNJIANG11,LEE12}. In particular, ML and WSF can consistently estimate \emph{arbitrarily close} DOAs given sufficient data and/or SNR. 

Cited sparse solvers introduce intrinsically a further \emph{DOA quantization} step and lose consistency, since the true array model is not in the solution space. However, DOA estimates on a fixed grid are extremely robust to noise and sufficiently close to the true values. If coarse DOA estimates are in the correct number and all within the trust region, off-grid DOA refinement can be performed optimally by classical local gradient or Newton descent ML or WSF optimization \cite{CADZOW90,VIBERG91}. Proposed extensions of sparse solvers to off-grid DOAs \cite{YANG13} cannot improve the performance of local ML and WSF estimators. In fact, sparse approaches cannot resolve multiple sources within each DOA bin and use a weak Taylor off-grid approximation of the array manifold. As shown in simulations, embedded off-grid DOA refinement at the earlier iterations is inefficient, since the large initial bias hampers convergence and increases the computational burden.

Prior estimation of the number of arriving paths is another essential information for ML DOA estimators. In a classical approach, the number of narrow-band paths is detected by analyzing the WSF error under different hypotheses by sequential F-test \cite{VIBERG91} or by Information Theoretic Criteria \cite{AKAIKE74,WAX85}. Such criteria are not well posed for quantized DOAs, since the true model is not among the choices \cite{AKAIKE74,BURNHAM04}. 

Following this idea, it is essential that the number of test hypotheses is minimized. In conclusion, the coarse DOA estimation stage (referred to as a \emph{booster}) for WSF has to find a possibly redundant number of candidate DOAs \cite{GERSHMAN99}, clustered around the true ones. At the same time, empty regions of the field of view must be excluded for accuracy reasons \cite{BUCRIS12} and effects of DOA quantization and covariance errors must be controlled.

From the previous discussion, any viable existing solution to initialization is iterative, suffers of intrinsic information loss by intermediate decision steps and exhibits often an irregular, non parallel flow graph. This is a severe issue in most practical applications of high resolution coherent DOA estimation in remote sensing (specular multipath, coherent jamming), seismics and wireless communications (radio-navigation, mobile devices), that need high accuracy and have critical \emph{latency}, rather than \emph{throughput} computational requirements in real time operations. 

Taking into account the described limitations, we developed a high resolution, \emph{strongly parallel} and \emph{non iterative} discrete DOA coarse estimator for general arrays, that is robust to low SNR and coherent source scenarios. This novel solution has been named Parallel BOOSTer (\emph{P-BOOST}).
 
P-BOOST circumvents the sidelobe issues intrinsic to the signal subspace formulation \cite{CADZOW90} of sparse DOA estimators and resorts to an original, information preserving combination of MUSIC, WSF and cross-validation concepts, valid even for coherent source scenarios. The overall number of operations is exchanged for a reduced processing latency in a parallel implementation \cite{KUNG85}.

In fact, MUSIC consistently estimates the DOAs and the spectral parameters of a (partially) uncorrelated point source \cite{SCHMIDT86} by \emph{inverting} in a constrained ML sense its steering vector from the sample signal subspace \cite{DICLAUDIO18}, asymptotically \emph{suppressing interference} from other sources. In a coherent scenario, the source wavefront is the \emph{weighted superposition} of the steering vectors of incoming paths and MUSIC cannot match successfully any single steering vector to the signal subspace \cite{SCHMIDT86}. 

However multi-dimensional \emph{Noise Subspace Fitting} (NSF) \cite{VIBERG91} approaches can be asymptotically efficient, but require a costly preliminary search. In this work we first look for the ML estimate of a linear combination of few steering vectors (referred to as \emph{composite steering vector}, CSV), mimicking multipath propagation \cite{SCHMIDT86}. We demonstrate that this estimator is feasible and is constituted by a \emph{multi-dimensional (MD-)MUSIC}, directly tied to the optimal NSF solution. Therefore, we propose to redefine the under-determined solution as a \emph{mixture of candidate CSVs}, each capable of super-resolution and approximated by a \emph{parallel and coarse} MD-MUSIC, starting at each candidate DOA and using a special noise subspace formulation of OMP. 

In Sect. \ref{sec:FOCUSSasMMSEbeamformingAndLinksToBayesianLearning} we show that energy localization is obtained in FOCUSS and SBS by imposing a sound amplitude prior which weights the codebook, according to a high resolution MMSE (Wiener) beamforming. This weighting cannot be approximated in the coherent case withoutseveral iterations. In fact, existing amplitude priors are mostly based on conventional beamforming, which is not optimal in any sense for the multi-source array model \cite{HARMANCI00}. 

In our approach, each CSV is first weighted by the power estimated by a conceptually close MV \emph{coherent beamformer} based on signal and noise sample subspaces. This is effective only because the CSVs obtained by MD-MUSIC are as close as possible to the signal subspace. Then, weighted CSVs are linearly combined by an \emph{approximate} ML MUSIC \cite{DICLAUDIO18} DOA estimator, which is a multi-dimensional generalization of previous weighted MUSIC approaches \cite{STOICA90a}. 

The output of this stage is used first to create a smoothed \emph{spatial (pseudo-)covariance estimate} based on a diagonal approximation of the source covariance, thus circumventing the coherency issue. This intermediate step still resembles FOCUSS and SBS resolution mechanisms. Finally, a high resolution source distribution is estimated by a \emph{cross-validation} approach, which is found equivalent to a \emph{Capon beamformer} \cite{LI05} applied to the weighted signal subspace.

It is shown that P-BOOST is more robust than conventional FOCUSS or OMP solutions while remaining computationally advantageous for its smaller overall processing latency. As regard the limit DOA resolution, P-BOOST shares the limitations of the other sparse solvers, due to DOA quantization \cite{BUCRIS12} and the MV beamforming structure, but retains the full  information for subsequent DOA refinements.

In fact, two main usages are foreseen for P-BOOST. The first application is to provide accurate information for the synthesis of \emph{array interpolation matrices} \cite{HUNG88,FRIEDLANDER93,WEISS95,DICLAUDIO05,BUCRIS12}, that map accurately the original array to virtual ULAs or other Vandermonde arrays within \emph{tight} source clusters. Array interpolation can exploit statistically and computationally efficient root-finding based DOA estimators, such as ROOT MUSIC \cite{RAO89} and MODE \cite{STOICA90a}, preserves Fisher information under mild conditions and can be made independent of the actual intra-cluster source configuration \cite{FRIEDLANDER93}. For instance, combining P-BOOST based interpolation and MODE yields a \emph{statistically efficient} and \emph{low latency} estimator for single DOA parameter search \cite{FRIEDLANDER93}.

The second foreseen P-BOOST application is WSF or ML boosting with \emph{general arrays} and \emph{multiple DOA parameters} for each source. In this case, intra-cluster DOA super-resolution is required right at the initialization stage. Still P-BOOST furnishes an improved prior source resolution for FOCUSS type estimators \cite{WIPF07}, followed by local WSF optimization. 

Detection with quantized DOA estimators is essential to reduce or avoid WSF hypothesis testing \cite{VIBERG91,GERSHMAN99}. To circumvent the DOA quantization issue, an approximate Akaike Information Criterion (AIC), based on the original formulation in the sense of the \emph{average likelihood over a set of experiments} \cite{AKAIKE74}, was developed for detecting the peaks of discrete spatial spectra, by simulating an ensemble of random DOA grids. This AIC was found equivalent to a sequential, regularized WSF over detected peaks and obtained excellent detection results over all the useful SNR range.

The paper is organized as follows. After a notation Sect. \ref{section:Notation} and a brief review of the narrow-band array model in Sect. \ref{sec:Arraysignalmodel}, the under-determined, sparse solutions are presented in Sect. \ref{sec:DiscretizedArrayModel} and re-interpreted as iterative Wiener beamforming in Sect. \ref{sec:FOCUSSasMMSEbeamformingAndLinksToBayesianLearning}. The non-iterative P-BOOST solution is detailed in Sect. \ref{sec:NonIterativeApproximateFOCUSSSolution} and analyzed for the computational complexity in Sect. \ref{sec:ComputationalAnalysis}. The approximate AIC source validation test is presented in Sect. \ref{sec:AICTypeSourceValidationFromSparseSpatialSpectra}. In Sect. \ref{sec:StatisticalValidationOfSparseSolutions} the problem of the statistical evaluation of sparse DOA estimators is critically afforded. In Sect. \ref{sec:ComputerSimulations}, P-BOOST is compared with existing approaches in computer simulations. Conclusion is drawn in Sect. \ref{sec:Conclusion}.

%-------------------------------------------------------------------------------------------------------------------------------------------------------------------------------------------------
\section{Notation}\label{section:Notation}
Throughout the paper matrices are indicated by boldface, capital letters, vectors by boldface, lowercase letters. The transpose of matrix $\bf A$ is indicated by ${\bf A}^T$, the Hermitian transpose by ${\bf A}^H$. ${{\bf{A}}^\dag }$ is the Moore-Penrose pseudoinverse of $\bf A$. ${\bf I}_M$ is the square identity matrix of size $M$. The operator ${\rm diag}\{{\bf A}\}$ creates a column vector with the main diagonal of $\bf A$, ${\rm diag}\{{\bf a}\}$ creates a diagonal matrix with the elements of vector ${\bf a}$ placed on its main diagonal. Sub-matrices are indexed by MATLAB-like conventions \cite{GOLUB89}. For instance, ${\bf A}(:,k)$ is the $k-$th column of matrix $\bf A$ and ${\bf A}(1:m,1:n)$ is the upper left submatrix of $\bf A$ of size $m\times n$. ${\rm trace}\{{\bf A}\}$ is the trace of $\bf A$. ${\rm det}\{{\bf A}\}$ is the determinant of $\bf A$. The Frobenius norm \cite{GOLUB89} of $\bf A$ is indicated by ${\left\| {\bf{A}} \right\|_F}$. The Hermitian square root ${{\bf{A}}^{{1 \mathord{\left/
 {\vphantom {1 2}} \right.
 \kern-\nulldelimiterspace} 2}}}$ of the positive semidefinite matrix $\bf A$ \cite{GOLUB89} obeys ${{\bf{A}}^{{1 \mathord{\left/
 {\vphantom {1 2}} \right.
 \kern-\nulldelimiterspace} 2}}} = {\left( {{{\bf{A}}^{{1 \mathord{\left/
 {\vphantom {1 2}} \right.
 \kern-\nulldelimiterspace} 2}}}} \right)^H}$ and ${{\bf{A}}^{{1 \mathord{\left/
 {\vphantom {1 2}} \right.
 \kern-\nulldelimiterspace} 2}}}{{\bf{A}}^{{1 \mathord{\left/
 {\vphantom {1 2}} \right.
 \kern-\nulldelimiterspace} 2}}} = {\bf{A}}$.

${\mathop{\rm E}\nolimits} \left\{ x \right\}$ is the expected value of the random variable $x$. Throughout the paper, empirical estimates are indicated by a \emph{hat superscript} (e.g., ${\hat{\bf A}}$ is the sample version of $\bf A$).

\section{Array signal model}
\label{sec:Arraysignalmodel}
A narrow-band array of $M$ sensors is immersed in a linear medium and receives $D$ signals $\left\{ {{s_1}\left( t \right),{s_2}\left( t \right), \ldots ,{s_D}\left( t \right)} \right\}$ impinging from $D$ directions, characterized by the parameter set ${\bf{\Theta }} = \left\{ {{{\bm{\theta }}_1},{{\bm{\theta }}_2}, \ldots ,{{\bm{\theta }}_D}} \right\}$, where the generic vector ${\bm{\theta }}_d$ contains the DOA parameters (azimuth, elevation, distance, etc..) of each arrival \cite{SCHMIDT86}. The $M$-dimensional array response vector (\emph{steering vector} \cite{SCHMIDT86}) ${\bf{a}}\left( {\bm{\theta }} \right)$ of each path impinging from the generic direction ${{\bm{\theta }}}$ is assumed known for the whole field of view of the array. 

The array model in discrete time is
\begin{equation}
	{\bf{x}}\left( n \right) = {\bf{A}}\left( {\bf{\Theta }} \right){\bf{s}}\left( n \right) + {\bf{v}}\left( n \right)
	\label{eqn:ArrayModel}
\end{equation}
for $n = 1, 2, \dots, N$, where the narrow-band \emph{snapshot} vector ${\bf{x}}\left( n \right)$ stacks the $M$ array output complex envelopes sampled at time $n$, ${\bf{A}}\left( {\bf{\Theta }} \right) = \left[ {\begin{array}{*{20}{c}}
{{\bf{a}}\left( {{{\bm{\theta }}_1}} \right)}&{{\bf{a}}\left( {{{\bm{\theta }}_2}} \right)}& \cdots &{{\bf{a}}\left( {{{\bm{\theta }}_D}} \right)}
\end{array}} \right]$ is the array transfer matrix of size $M \times D$. ${\bf{s}}\left( n \right) = {\left[ {\begin{array}{*{20}{c}}
 {{s_1}\left( {n} \right)}&{{s_2}\left( {n} \right)}& \cdots &{{s_D}\left( {n} \right)}
\end{array}} \right]^T}$, of size $D \times 1$, is the vector of the signals of interest at discrete time $n$ and ${\bf{v}}\left( n \right)$ is the additive noise vector of length $M$.

Signal and noise are assumed as realizations of zero mean random circular Gaussian processes\footnote{Straight extensions are possible for signal and noise distributions having finite fourth-order moments \cite{HOFFBECK96,LEDOIT04}.}  The noise is assumed statistically independent of the signals, with known covariance ${{\bf{R}}_{vv}} = E\left\{ {{\bf{v}}\left( n \right){{\bf{v}}^H}\left( n \right)} \right\}$, except for a positive scalar $\lambda_v$. 

The spatial covariance matrix (SCM) ${{\bf{R}}_{xx}} = E\left\{ {{\bf{x}}\left( n \right){{\bf{x}}^H}\left( n \right)} \right\}$ is modeled as
\begin{equation}
{{\bf{R}}_{xx}} = {\bf{A}}\left( {\bf{\Theta }} \right){\bf{P}}{\bf{A}}{\left( {\bf{\Theta }} \right)^H} + {\lambda _v}{{\bf{R}}_{vv}}
\label{eqn:SCM}
\end{equation}
where ${\bf{P}} = E\left\{ {{\bf{s}}\left( n \right){{\bf{s}}^H}\left( n \right)} \right\}$ is the signal covariance matrix of size $D$ and rank $K \le D$.

For the non ambiguity of the model, it is assumed that $D<M$ and that any set of $D$ steering vectors are linearly independent, at least in a neighborhood of the true DOAs. Further limitations on $D$ apply for coherent signals, i.e., for $K<D$ and/or multi-parameter ${\bm{\theta }}$ \cite{SCHMIDT86,VIBERG91}.

The eigen-decomposition (EVD) of the noise whitened covariance matrix ${{\bf{S}}_{xx}} = {\bf{R}}_{vv}^{-{1 \mathord{\left/
 {\vphantom {1 2}} \right.
 \kern-\nulldelimiterspace} 2}}{{\bf{R}}_{xx}}{\bf{R}}_{vv}^{-{1 \mathord{\left/
 {\vphantom {1 2}} \right.
 \kern-\nulldelimiterspace} 2}}$ is \cite{SCHMIDT86}
\begin{equation}
	\begin{array}{*{20}{c}}
{{{\bf{S}}_{xx}} = {\bf{R}}_{vv}^{ - {1 \mathord{\left/
 {\vphantom {1 2}} \right.
 \kern-\nulldelimiterspace} 2}}{\bf{A}}\left( {\bm{\Theta }} \right){\bf{P}}{\bf{A}}{{\left( {\bm{\Theta }} \right)}^H}{\bf{R}}_{vv}^{ - {1 \mathord{\left/
 {\vphantom {1 2}} \right.
 \kern-\nulldelimiterspace} 2}}} + \lambda_v {\bf I}_M \\
{ = {{\bf{E}}_s}{{\bf{\Lambda }}_s}{\bf{E}}_s^H + {\lambda _v}{{\bf{E}}_v}{\bf{E}}_v^H}
\end{array}
\label{eqn:CovarianceEVD}
\end{equation}
where ${{\bf{E}}_s}$ is the orthonormal \emph{signal subspace} basis corresponding to the $K$ dominant eigenvalues ${\lambda _1} \ge {\lambda _2} \ge \ldots \ge {\lambda _K} > {\lambda _v}$, ${{\bf{\Lambda }}_s} = {\mathop{\rm diag}\nolimits} \left \{ {\left[ {\begin{array}{*{20}{c}}
{{\lambda _1}}& \cdots &{{\lambda _K}}
\end{array}} \right]} \right\}$ and ${{\bf{E}}_v}$ is the orthogonal complement of ${{\bf{E}}_s}$ \cite{GOLUB89} and defines the \emph{noise subspace} basis.
 
\section{Spatially Discretized Array Model}
\label{sec:DiscretizedArrayModel}
Several approaches exist for consistently estimating the free parameters of \eqref{eqn:SCM}, namely ${\bf{P}}$, ${\bf{\Theta }}$ and $\lambda_v$, from a consistent estimate ${\hat{\bf{R}}_{xx}}$ of ${{\bf{R}}_{xx}}$ obtained from $N > M$ snapshots. In particular, ML \cite{BOHME85}, WSF and NSF \cite{VIBERG91} are asymptotically efficient for $N/M \rightarrow\infty$ and Gaussian signals and noise. However, they have several local minima, so that they require a multidimensional search, initialized by a suboptimal estimator within a \emph{fraction} of the array beam-width and an \emph{exact} knowledge of the number $D$ of paths\footnote{The number $K$ of independent signals, equal to the covariance rank, can be estimated by Information Theoretic Criteria \cite{WAX85,NADAKUDITI08} and it is not overly critical, at least for WSF.}.

The inability of handling coherent sources \cite{SCHMIDT86,ROY89} and the worse low SNR estimation threshold with respect to ML and WSF \cite{MESTRE08} of MUSIC and ESPRIT, as well as the limited resolution of classical beamforming \cite{VANTREES02}, are show stoppers for preliminary DOA estimation in such difficult cases. 

In the sequel, we will concentrate on the WSF equation
\begin{equation}
{{{\hat{\bm \Theta }}}_{{\rm{WSF}}}} = \mathop {\arg \min }\limits_{{\bm{\Theta }},{{\bf{S}}_{{\rm{WSF}}}}} \left\| {{{\hat {\bf{E}}}_s}{{\hat {\bf{W}}}_s} - {{\bf{R}}_{vv}^{ - {1 \mathord{\left/
 {\vphantom {1 2}} \right.
 \kern-\nulldelimiterspace} 2}}}{\bf{A}}\left( {\bm{\Theta }} \right){{\bf{S}}_{{\rm{WSF}}}}} \right\|_F^2
\label{eqn:WSF}
\end{equation}
where the \emph{empirical signal subspace} ${{{\hat{\bf E}}}_s}$ and the \emph{empirical subspace weighting }matrix ${\hat{\bf W}}_s$ of size $K \times K$ are derived from the EVD of the empirical whitened covariance ${\hat {\bf{S}}_{xx}} = {\bf{R}}_{vv}^{ - {1 \mathord{\left/
 {\vphantom {1 2}} \right.
 \kern-\nulldelimiterspace} 2}}{{{\hat{\bf R}}}_{xx}}{\bf{R}}_{vv}^{ - {1 \mathord{\left/
 {\vphantom {1 2}} \right.
 \kern-\nulldelimiterspace} 2}}$, with ${\hat \lambda _v} = {\left( {M - K} \right)^{ - 1}}\sum\limits_{k = K + 1}^M {{{\hat \lambda }_k}} $. ${\bf{S}}_{\rm{WSF}}$ is a complex mixing matrix of size $D \times K$ \cite{VIBERG91}. 

The use of empirical subspaces is not a limitation for \eqref{eqn:WSF} even in a small $N$ scenario. First, SCM eigenvectors and eigenvalues map one-to-one the snapshot information relevant for unconditional (i.e., deterministic, but unknown signals) or conditional (random signals) Gaussian assumptions \cite{DORON93}. Second, $N = K+1$ snapshots are the minimum required for an independent noise variance and DOA parameter identification. Third, \eqref{eqn:WSF} is equivalent to an unconditional ML estimator \cite{DORON93}, but based on a \emph{regularized SCM estimate} \eqref{eqn:CovarianceEVD} with equalized noise eigenvalues. Regularization lowers the MSE w.r.t. the original SCM estimate \cite{LEDOIT04}. More elaborate SCM regularization techniques are effective for small samples, non Gaussian, long-tailed signals and outliers \cite{HOFFBECK96,LEDOIT04}.

The relevance of WSF lies in the general statistical characterization of the fitting error of \eqref{eqn:WSF} \cite{VIBERG91}. In fact, the WSF residuals, originated by the spurious weighted projections of the sample ${{\hat{\bf E}}}_s$ onto the true noise subspace ${\bf E}_v$, for $N \gg M$ asymptotically approach a $\left( {M - D} \right) \times K$ Gaussian \emph{random matrix} with i.i.d. circular entries with zero mean and ${N^{ - 1}}$ variance for the optimal signal eigenvector weighting\footnote{The choice ${{\bf{W}}_{{s}}} = {\rm{diag}}\left\{ {\sqrt {\frac{{{{\left( {{\lambda _k} - {\lambda _v}} \right)}^2}}}{{{\lambda _k}{\lambda _v}}} + \frac{{M - K}}{N}} ;k = 1, \ldots ,K} \right\}$ was justified for finite sample in \cite{DICLAUDIO18} and adopted in this work. The empirical estimate ${\hat{\bf W}}_s$ of ${{\bf W}_s}$ is used in \eqref{eqn:WSF}. } \cite{VIBERG91} 
\begin{equation}
{{\bf W}_s} = {\mathop{\rm diag}\nolimits} \left\{ {\frac{{{\lambda _k} - {\lambda _v}}}{{\sqrt {{\lambda _k}{\lambda _v}} }};k = 1, \ldots ,K} \right\} \;.
\label{eqn:WSFOptimalWeights}
\end{equation}

This asymptotic distribution constitutes a useful support in a Gaussian scenario, but WSF and NSF approximate locally the \emph{Best Linear Unbiased Estimate} (BLUE) for DOAs even for non-Gaussian, relatively small sample subspace errors \cite{VIBERG91,KAY93}. Another advantage of WSF w.r.t. the conditional ML estimator \cite{BOHME85} is that its cost function vanishes for $N/M \rightarrow \infty$ at the true DOAs and flattens out when the fitting order exceeds $D$ \cite{VIBERG91,GERSHMAN99}.

After estimating $K$ by Information Theoretic criteria \cite{AKAIKE74,WAX85,NADAKUDITI08}, the number $D$ of paths can be estimated by running WSF or MODE with increasing $D \ge K$ and statistically checking the fitting error for the multiple hypotheses tested \cite{VIBERG91,GERSHMAN99}. 

The choice of the initial angles is the most critical part of \eqref{eqn:WSF}. Suboptimal parametric DOA estimators cannot reliably provide them at low SNR or with coherent sources. For ULAs, spatial smoothing \cite{WAX94} or its evolved iterative version MODE \cite{STOICA90a} actually works, but the first iteration is critical for convergence at low SNR because of ill-conditioning of internal equations \cite{GERSHMAN99,DICLAUDIO18}.

Under-determined solvers like FOCUSS \cite{GORODNITSKY97} are attractive for coarse DOA estimation, being sparse solutions of the spatially discretized integral equation
\begin{equation}
	\int\limits_{FOV\left( {\bm{\theta }} \right)} {{\bf{R}}_{vv}^{-{1 \mathord{\left/
 {\vphantom {1 2}} \right.
 \kern-\nulldelimiterspace} 2}}{\bf{a}}\left( {\bm{\theta }} \right){\bf{S}}\left( {\bm{\theta }} \right)d{\bm{\theta }}} \approx {{{\hat{\bf E}}}_s}{{{\hat{\bf W}}}_s} 
	\label{eqn:IntegralEquation}
\end{equation}
where ${\bf{S}}\left( {\bm{\theta }} \right) = \left[ {\begin{array}{*{20}{c}}
{{S_1}\left( {\bm{\theta }} \right)}& \cdots &{{S_K}\left( {\bm{\theta }} \right)}
\end{array}} \right]$ is a row vector of point mass functions of the kind ${S_k}\left( {\bm{\theta }} \right) = \sum\limits_{d = 1}^D {{s_{k,d}}\delta \left( {{\bm{\theta }} - {{\bm{\theta }}_d}} \right)} $ for $k=1,2,\ldots,K$, where $\delta \left( {{\bm{\theta }} - {{\bm{\theta }}_d}} \right)$ is the (multi-dimensional) Dirac pulse at ${{\bm{\theta }} = {{\bm{\theta }}_d}}$.

In fact, finite element discretization of \eqref{eqn:WSF} leads to the under-determined equation set
\begin{equation}
	{\bf{BS}} \simeq {{\hat{\bf E}}_s}{{\hat{\bf W}}_s} + {{\hat{\bf E}}_v}{\bf{G}}
		\label{eqn:SparseSystem}
\end{equation}
where ${\bf{B}}\left( {:,q} \right) = {\bf{R}}_{vv}^{-{1 \mathord{\left/
 {\vphantom {1 2}} \right.
 \kern-\nulldelimiterspace} 2}}{\bf{a}}\left( {{{\bm{\theta }}_q}} \right)$ for $q=1,2,\ldots,Q \gg M$ is the \emph{codebook} matrix, $\bf S$ is a $Q \times K$ unknown matrix, whose support is ideally sparse \cite{GORODNITSKY97,DONOHO03,WIPF07} and \emph{covers the entire field of view}. ${\bf{G}}$ is the random WSF error matrix of size $(M-K) \times K$ with zero mean, i.i.d., circular entries of variance $N^{-1}$ \cite{VIBERG91,DICLAUDIO18}. The sources are located by searching for the non-zero rows of $\bf S$.

Unfortunately, by the assumptions made in Sect. \ref{sec:Arraysignalmodel}, the representation of a source with DOA not coincident with one codebook ${\bm{\theta }}_q$ would require in general $M$ non zero entries in each column of $\bf S$. In addition, each $M \times M$ submatrix of $\bf B$ tends to have infinite condition number for $\inf \left\| {{{\bm{\theta }}_q} - {{\bm{\theta }}_l}} \right\| \to 0$ and therefore the codebook does not obey the Restricted Isometry Property \cite{CANDES05} for an unique solution that would be biased anyway. Finally, reducing DOA quantization leads to diverging memory and computational requirements. 

In the sequel we will mainly refer to the FOCUSS family \cite{GORODNITSKY97}, since we verified that it performed generally better than $L_1$ penalized fitting \cite{MALIOUTOV05} on empirical data. In addition, FOCUSS and related SBS solvers \cite{WIPF07} are interpreted as a bank of \emph{MMSE (Wiener) beamformers} \cite{DEMOOR93}, closely related to Capon beamformers \cite{LI05} and depending on prior information about source amplitudes. We show that the MUSIC paradigm can optimally furnish this information if revamped as the constrained ML estimator of a proper \emph{linear combination} of steering vectors.
 
\section{FOCUSS as MMSE beamforming and links to Bayesian Learning}
\label{sec:FOCUSSasMMSEbeamformingAndLinksToBayesianLearning}
The basic FOCUSS solution \cite{GORODNITSKY97}\footnote{A dual form exists \cite{WIPF07}, but it is less useful for our purposes.} can be written as
\begin{equation}
	\begin{array}{c}
{{{\hat{\bf S}}}_\mathrm{FOCUSS}} = \mathop {\arg \min }\limits_{\bf{S}} \left\{ {\sum\limits_{q = 1}^Q {\ln } \left\| {{\bf{S}}\left( {q,:} \right)} \right\|_2^2} \right.\\
\left. { + \lambda \left\| {{{\hat {\bf{E}}}_s}{{\hat {\bf{W}}}_s} - {\bf{BS}}} \right\|_F^2} \right\}
\end{array}
	\label{eqn:FOCUSS}
\end{equation}
and it is iteratively estimated. At the equilibrium, setting the gradient of \eqref{eqn:FOCUSS} to zero leads to
\begin{equation}
	\left( {{\bf{D}}_S^{ - 2} + \lambda {{\bf{B}}^H}{\bf{B}}} \right){\hat{\bf S}}_\mathrm{FOCUSS} = \lambda {{\bf{B}}^H}{\hat {\bf{E}}_s}{\hat {\bf{W}}_s}
	\label{eqn:FOCUSSgradient}
\end{equation}
where ${{\bf{D}}_S} = {\rm{ }}{K^{ - {1 \mathord{\left/
 {\vphantom {1 2}} \right.
 \kern-\nulldelimiterspace} 2}}}{\mathop{\rm diag}\nolimits} \left\{ {{{\left\| {{{{\hat{\bf S}}}_\mathrm{FOCUSS}}\left( {q,:} \right)} \right\|}_2}} \right\}$ and ${\bf{D}}_S^{ - 1}$ is intended as pseudoinverse. Error vanishes for $\lambda \to \infty$ \cite{GOLUB89}. Replacing ${\bf{Y}} = {\bf{D}}_S^{ - 1}{\hat{\bf S}}_\mathrm{FOCUSS}$ and multiplying both members of \eqref{eqn:FOCUSSgradient} by ${\bf{D}}_S$ shows that ${\bf{Y}}$ is the minimum $L_2$ norm solution to the under-determined linear system
\begin{equation}
\left( {{\bf{B}}{{\bf{D}}_S}} \right){\bf{Y}} = {{{\hat{\bf E}}}_s}{{{\hat{\bf W}}}_s} 
	\label{eqn:WeightedFOCUSS}
\end{equation}
where ${\left\| {{\bf{Y}}\left( {q,:} \right)} \right\|_2^2} = K$ or zero. 

A very interesting point is that this solution coincides with the \emph{MMSE beamforming estimate} of $\bf S$  
	\begin{equation}
	{{{\hat{\bf S}}}_{{\rm{FOCUSS}}}} = {\left[ {{{\left( {{\bf{BD}}_S^2{{\bf{B}}^H}} \right)}^{ - 1}}{\bf{BD}}_S^2} \right]^H}{\hat{\bf E}}_s{{{\hat{\bf W}}}_s} 
			\label{eqn:MMSEbeamformer}
	\end{equation}
assuming $E\left\{ {{\bf{Y}}{{\bf{Y}}^H}} \right\} = K{{\bf{I}}_Q}$ \cite{KAY93}. In particular, the $M \times M$ matrix 
\begin{equation}
	{{\bf{R}}_{ss}} = {\bf{BD}}_S^2{{\bf{B}}^H}
	\label{eqn:Rss}
\end{equation}
plays the role of a noiseless array \emph{pseudo-covariance} for uncorrelated sources, which cancels the sidelobes of the beamformer ${{{{\bf{R}}_{ss}^{ - 1}}}{\bf{BD}}_S^2}$ even in a fully coherent scenario.
	
This property suggests that the convergence of FOCUSS to a sparse solution is tied to imposing the \emph{appropriate amplitude prior} ${\bf{D}}_S$, enhancing the manifold around the true DOAs. In \cite{WIPF07} the problem was systematically afforded in a Bayesian framework, leading to several iterative generalizations (SBSs) of FOCUSS. In particular, the Maximum a Posteriori (MAP) and SBS iterations for \eqref{eqn:WSF} and zero mean Gaussian prior amplitudes with diagonal covariance ${\bf{D}}_S^2$ are identical \cite{KAY93} and expressed by
\begin{equation}
	{\hat{\bf S}}_\mathrm{MAP} = {{\bf{D}}_S}{\left( {{{\bf{D}}_S}{{\bf{B}}^H}{\bf{B}}{{\bf{D}}_S} + {N^{ - 1}}{{\bf{I}}_Q}} \right)^{ - 1}}{{\bf{D}}_S}{{\bf{B}}^H}{{{\hat{\bf E}}}_s}{{{\hat{\bf W}}}_s}\;.
	\label{eqn:Gaussian SBS}
\end{equation}

By inserting the reduced size SVD \cite{GOLUB89} ${{\bf{D}}_S}{{\bf{B}}^H} = {{\bf{U}}_B}{{\bf{\Sigma }}_B}{\bf{V}}_B^H$, it is shown that \eqref{eqn:Gaussian SBS} is a version of \eqref{eqn:FOCUSS} with finite $\lambda$, which leads to
\begin{equation}
	{{\hat{\bf S}}_{{\rm{MAP}}}} = {{\bf{D}}_S}{{\bf{U}}_B}{{\bf{\Sigma }}_B}{\left( {{\bf{\Sigma }}_B^2 + {N^{ - 1}}{{\bf{I}}_M}} \right)^{ - 1}}{\bf{V}}_B^H{{\hat{\bf E}}_s}{{\hat{\bf W}}_s}\;.
	\label{eqn:MAP_SVD}
\end{equation}

This equation reduces for $N\rightarrow\infty$ to the classical FOCUSS estimate \eqref{eqn:MMSEbeamformer}, rewritten with the same notation as
\begin{equation}
	{\hat{\bf S}}_\mathrm{FOCUSS} = {{\bf{D}}_S}{\bf{U}}_B{{\bf{\Sigma }}_B^{ - 1}}{{\bf{V}}_B^H}{{{\hat{\bf E}}}_s}{{{\hat{\bf W}}}_s} \; .
\label{eqn:FOCUSS_SVD}
\end{equation}
 
In existing forms of SBS the initial amplitude distribution ${\bf{D}}_S$ is given by conventional beamforming \cite{CADZOW90,SELVA17} with the cited problems of sidelobes, spurious sources and coherent interference patterns. In addition, the strongly weighted codebook ${\bf{BD}}_S$ of \eqref{eqn:WeightedFOCUSS} amplifies array manifold errors in all \emph{signal subspace approaches}, increasing bias \cite{SWINDLEHURST93,BUCRIS12}. 

On the contrary, MUSIC is notoriously robust to steering vector errors and largely insensitive to sidelobes, because of the implicit source power pre-whitening \cite{SWINDLEHURST92,LEE12}. For this reason, a sparse approach based on noise subspace is herein proposed as a logical alternative, following the hints of \cite{FANNJIANG11}.

\section{Non Iterative MUSIC Based Parallel Booster (P-BOOST)}
\label{sec:NonIterativeApproximateFOCUSSSolution}
The spectral MUSIC \cite{SCHMIDT86} is suboptimal \cite{STOICA90a}. However asymptotically efficient DOA estimators based on the noise subspace do exist, such as NSF and MODE \cite{STOICA90a}. In the sequel, we will show that spectral MUSIC limitations can be circumvented by working on suitable \emph{linear combinations} of few codebook steering vectors, herein referred to as CSVs. This fact can be put into evidence by re-writing NSF for our purposes. Equation \eqref{eqn:SparseSystem} is projected onto consistent estimates ${{\hat{\bf E}}_s}$ and ${{\hat{\bf E}}_v}$ of signal and noise subspaces \cite{DICLAUDIO18}
\begin{equation}
	{\hat{\bf W}}_s^{ - 1}{\hat{\bf E}}_s^H{\bf{BS}} \simeq {{\bf I}_K} 
\label{eqn:ML_MUSIC_signal}
\end{equation}
\begin{equation}
	{\hat{\bf E}}_v^H{\bf{BS}} \simeq {\bf{G}} \;.
\label{eqn:ML_MUSIC_noise}
\end{equation}

The pair \eqref{eqn:ML_MUSIC_signal} and \eqref{eqn:ML_MUSIC_noise} with additional $o\left( {{N^{ - 1/2}}} \right)$ terms was exploited to develop a \emph{ML MUSIC} subspace estimator \cite{DICLAUDIO18}. However, the classical NSF solves only \eqref{eqn:ML_MUSIC_noise}, assuming a Gaussian $\bf G$ \cite{VIBERG91} and minimizing the negative log-likelihood
\begin{equation}
\begin{array}{rcl}
	{\cal{L}_\mathrm{NSF}} & = & (M - K)\left[ {K\ln \left( {{\pi  \mathord{\left/
 {\vphantom {\pi  N}} \right.
 \kern-\nulldelimiterspace} N}} \right) - \ln \det \left( {{{\bf{S}}^H}{\bf{S}}} \right)} \right] \\
& & + N\left\| {{\hat{\bf E}}_v^H{\bf{BS}}} \right\|_F^2
\end{array}
\label{eqn:NSF_likelihood}
\end{equation}
w.r.t. the DOAs and the sparse mixing matrix $\bf S$ of size $Q \times K$ (up to a unitary right transformation). Each column of $\bf S$ defines a CSV with at most $D$ nonzero coefficients. In comparison with \eqref{eqn:NSF_likelihood}, spectral MUSIC \cite{SCHMIDT86} is sub-optimal, since it tries to \emph{recover a single steering vector} ${\bf{B}}\left( {:,q} \right)$ from ${\bf{BS}}$. This is not possible for coherent scenarios, where $\bf S$ is not full rank.

The basic idea pursued in this work stems from observing that if a certain ${\bf{B}}\left( {:,q} \right)$ is in the WSF \eqref{eqn:WSF} or NSF solution \eqref{eqn:NSF_likelihood}, it is possible to right multiply $\bf S$ by an unitary matrix of size $K$, so that all the coefficients referred to the $q$-th angle of the CSVs are zero except one that may be nonzero. Specifically, it is sufficient that one column of the unitary matrix is ${{{{\bf{S}}{{\left( {q,:} \right)}^H}} \mathord{\left/
 {\vphantom {{{\bf{S}}{{\left( {q,:} \right)}^H}} {\left\| {{\bf{S}}\left( {q,:} \right)} \right\|}}} \right.
 \kern-\nulldelimiterspace} {\left\| {{\bf{S}}\left( {q,:} \right)} \right\|}}_2}$. However, one CSV may not contain significant terms corresponding to all the DOAs and the coefficients of the vast majority of codebook steering vectors are negligible.

Then, a ML estimator can be applied to this particular CSV, searching across multiple candidate DOAs. As shown in the sequel, this procedure implements a sort of \emph{multi-dimensional MUSIC} (MD-MUSIC). The goal is to obtain \emph{in parallel} a set of $Q$ \emph{coarse} and sparse CSVs \eqref{eqn:NSF_likelihood}, each starting from one codebook DOA. The array response is finally reconstructed as an under-determined linear combination of these CSVs by an \emph{approximate} ML MUSIC approach \cite{DICLAUDIO18}, combining \eqref{eqn:ML_MUSIC_signal} and \eqref{eqn:ML_MUSIC_noise}. The reconstruction is not yet sparse, because finite sample errors and angle quantization induce a different DOA angular spread in each CSV. Nevertheless it is strongly energy localized near the true DOAs \cite{GORODNITSKY97}, because of the high robustness of MD-MUSIC to sample finiteness and angular uncertainties. In fact, MD-MUSIC is a constrained ML solution retaining the low sensitivity of MUSIC to manifold errors \cite{SWINDLEHURST92}.

The ML MUSIC solution will be the basis for a robust estimate of a diagonally smoothed pseudo-covariance ${{\bf{R}}_{ss}}$, similar to \eqref{eqn:Rss}, used for estimating source amplitudes through a \emph{cross-validation} problem based on \eqref{eqn:SparseSystem} and insensitive to coherency effects. These points are detailed in the following sections.

\subsection{OMP MD-MUSIC Parallel Pre-Processing}
\label{sec:OMPMDMUSIC}
The CSV minimizing \eqref{eqn:NSF_likelihood} referred to the $q$-th candidate DOA is readily obtained by minimizing the variance
\begin{equation}
\sigma _q^2 = \left\| {{\bf{S}}\left( {:,q} \right)} \right\|_2^{ - 2} = \frac{N}{{M - K}}\left\| {\hat {\bf{E}}_v^H{{{\hat{\bf b}}}_q}} \right\|_2^2
	\label{eqn:md-music}
\end{equation}
for a constrained CSV defined as
\begin{equation}
		{\hat{\bf b}}_q = {\bf{B}}{{\bf{s}}_q} 
		\label{eqn:MD-MUSICslice}
	\end{equation}
where ${\bf{s}}_q$ is a sparse $Q \times 1$ coefficient vector having $D_q \le M-K-1$ nonzero complex valued coefficients and ${{\bf{s}}_q}\left( q \right) \ne 0$. To avoid ill-conditioning problems when ${\bf{B}}\left( {:,q} \right)$ is not in the solution of \eqref{eqn:MD-MUSICslice} (i.e., ${{\bf{s}}_q}\left( q \right) \approx 0$), ${\hat{\bf b}}_q$ is rewritten \emph{one-to-one} as the original ${\bf{B}}\left( {:,q} \right)$, plus an orthogonal, but DOA annotated component
\begin{equation}
	{\hat{\bf b}}_q = {\bf{B}}\left( {:,q} \right) + {{\bf{\Pi }}_k}{\bf{B}}\left( {:,\left[ {1:q - 1,q + 1:Q} \right]} \right){{\bf{c}}_q} \;
	\label{eqn:new_codebook_vector}
\end{equation}
for $q=1,2,\ldots,Q$, where
\begin{displaymath}
	{{\bf{\Pi }}_k} = {{\bf{I}}_M} - \frac{{{\bf{{B}}}(:,q){\bf{{B}}}{{(:,q)}^H}}}{{\left\| {{\bf{{B}}}(:,q)} \right\|_2^2}}
\end{displaymath}
and ${{\bf{c}}_q}$ is an auxiliary sparse $(Q-1) \times 1$ coefficient vector. ${{\bf{s}}_q}$ is obtained by equating \eqref{eqn:MD-MUSICslice} and \eqref{eqn:new_codebook_vector}. 

In order to improve over the classical MUSIC peak picking detection, OMP \cite{CADZOW90,MALLAT93,LEE12} is used for finding up to $M-K-1$ candidate DOAs by sequentially minimizing \eqref{eqn:md-music}. In fact, OMP is nearly ideal for this task because of the good conditioning of \eqref{eqn:md-music} and the strong sidelobe suppression of NSF approaches \cite{MALLAT93}. 

OMP yields coarse DOA estimates, but with a scatter around the true DOAs among different CSVs due to grid mismatching and finite sample errors. If $M-K=1$ at most $M-1$ partially uncorrelated sources can be identified and the MD-MUSIC corresponds to ordinary MUSIC \cite{SCHMIDT86}, otherwise \eqref{eqn:new_codebook_vector} can find candidate coherent DOAs far from the reference DOA.

\subsection{Coherent ML MUSIC estimate}
\label{sec:CoherentMLMUSICEstimate}
In our approach, the vectors ${\hat{\bf b}}_q$ for $q,1,2,\ldots,Q$ obtained from MD-MUSIC can constitute a valid basis for the solution of \eqref{eqn:SparseSystem}. However, the fitting problem \eqref{eqn:SparseSystem} remains under-determined and, at the same time, we seek for a well-approximate, non-iterative solution. By \eqref{eqn:WeightedFOCUSS} we know that the analytic minimum norm solution of \eqref{eqn:SparseSystem} for $\bf S$ is optimal in the MMSE sense only for $E\left\{ {{\bf{S}}{{\bf{S}}^H}} \right\} \propto {{\bf{I}}_Q}$ \cite{KAY93}. In particular, \eqref{eqn:WeightedFOCUSS} shows that an energy localized solution $\bf S$ approaching MMSE optimality can be obtained employing a \emph{whitening transformation} ${\bf{Y}} = {\bf{R}}_{ss}^{-{1 \mathord{\left/
 {\vphantom {1 2}} \right.
 \kern-\nulldelimiterspace} 2}}{\bf{S}}$ and solving for the minimum ${\left\| {\bf{Y}} \right\|_F}$. Differently from \eqref{eqn:Rss}, a \emph{coherent pseudo-covariance} ${{\bf{R}}_{ss}}$ is sought in the form
\begin{equation}
{{\bf{R}}_{ss}} = \sum\limits_{q = 1}^Q {\beta _q^2{{\hat {\bf{b}}}_q}\hat {\bf{b}}_q^H}
\label{eqn:CoherentPseudoCovariance}
\end{equation}
where the ${\beta_q} \ge 0$ are unknown scaling coefficients to be optimized. The various ${\hat{\bf b}}_q$ are in general mutually strongly correlated from \eqref{eqn:NSF_likelihood}, but, as in FOCUSS \eqref{eqn:WeightedFOCUSS}, they are initially \emph{assumed} as independent.

To avoid iterations, inspired by the MMSE beamformer interpretation \eqref{eqn:MMSEbeamformer} of FOCUSS, in the absence of prior information the weights ${\beta_q}$ are estimated by first solving the MV beamforming problem
\begin{equation}
	{{\bf{h}}_q} = \mathop {\arg \min }\limits_{\bf{h}} \left( {{{\bf{h}}^H}E\left\{ {{{{\hat{\bf E}}}_s}{\hat{\bf W}}_s^2{\hat{\bf E}}_s^H} \right\}{\bf{h}}} \right)
\label{eqn:HqEstimate}	
\end{equation}
with the distortion-less constraint ${\hat{\bf b}}_q^H{\bf{h}}_q = 1$ . Under the model \eqref{eqn:SparseSystem}\footnote{More complex solutions are possible for this regularized estimate based on \cite{LEDOIT04,LI05}.} we can set $E\left\{ {{{{\hat{\bf E}}}_s}{\hat{\bf W}}_s^2{\hat{\bf E}}_s^H} \right\} \approx {{{\hat{\bf E}}}_s}{\hat{\bf W}}_s^2{\hat{\bf E}}_s^H + K{N^{ - 1}}{{{\hat{\bf E}}}_v}{\hat{\bf E}}_v^H$, which, after straight algebra, leads to
\begin{equation}
	\begin{array}{c}
\beta _q^2 = {\bf{h}}_q^H{{{\hat{\bf E}}}_s}{\hat{\bf W}}_s^2{\hat{\bf E}}_s^H{{\bf{h}}_q} \\
= \displaystyle\frac{{{\hat{\bf b}}_q^H{{{\hat{\bf E}}}_s}{\hat{\bf W}}_s^{ - 2}{\hat{\bf E}}_s^H{{{\hat{\bf b}}}_q}}}{{{{\left[ {{\hat{\bf b}}_q^H\left( {{{{\hat{\bf E}}}_s}{\hat{\bf W}}_s^{ - 2}{\hat{\bf E}}_s^H + N{K^{ - 1}}{{{\hat{\bf E}}}_v}{\hat{\bf E}}_v^H} \right){{{\hat{\bf b}}}_q}} \right]}^2}}} \;.
\end{array}
\label{eqn:SquaredBeta}
\end{equation}
 
A keynote is that $\beta _q^2$ is significant only when ${\hat{\bf b}}_q$ is close to the signal subspace, otherwise source cancellation occurs \cite{LI05}. As a consequence, applying \eqref{eqn:SquaredBeta} to the original codebook in a coherent case leads to very small $\beta _q^2$s even along the true DOAs. 

Up to an arbitrary scaling and an inessential rotation of ${\bf{Y}}$, the overall whitening transformation defined by \eqref{eqn:CoherentPseudoCovariance} and \eqref{eqn:SquaredBeta} can be chosen as
\begin{equation}
{\bf{BS}} = {\bf{B}}\hat{\bf{C}}{\bf{Y}} = \left[ {\begin{array}{*{20}{c}}
{{\beta _1}{{\hat {\bf{b}}}_1}}& \cdots &{{\beta _Q}{{\hat {\bf{b}}}_Q}}
\end{array}} \right]{\bf{Y}}
\label{eqn:coherent_solution}
\end{equation}
where $\hat{\bf{C}} = \left[ {\begin{array}{*{20}{c}}
{{\beta _1}{{{\bf{s}}}_1}}& \cdots &{{\beta _Q}{{{\bf{s}}}_Q}}
\end{array}} \right]$ and ${\bf{S}} = \hat{\bf{C}}{\bf{Y}}$.

Replacing \eqref{eqn:coherent_solution} into \eqref{eqn:SparseSystem}, right multiplying both members by a $K \times Q$ unknown weight matrix ${{\bf{Z}}}$ and neglecting $o\left( {{N^{ - 1/2}}} \right)$ terms \cite{DICLAUDIO18} leads to the under-determined Paige type equation set \cite{GOLUB89}
\begin{equation}
	{\hat {\bf{E}}_s}{\hat {\bf{W}}_s}{\bf{Z}} + {\hat {\bf{E}}_v}{\bf{GZ}} = {\bf{B}}{\hat{\bf C}}{\bf{Y}} \;.
	\label{eqn:ML-MUSIC-mat}
\end{equation}

Under the Gaussian assumption on the random matrix ${\bf{G}}$ given in \eqref{eqn:SparseSystem} \cite{STOICA90a,VIBERG91}, \eqref{eqn:ML-MUSIC-mat} can be solved in a ML MUSIC sense\cite{DICLAUDIO18}. 

The solution for ${\bf{Z}}$ conditioned to ${\bf{Y}}$ is obtained immediately by projecting both members of \eqref{eqn:ML-MUSIC-mat} onto ${\hat {\bf{E}}_s}$ and is ${\bf{Z}} = \hat {\bf{W}}_s^{ - 1}\hat {\bf{E}}_s^H {\bf{B}}{\hat{\bf C}}{\bf{Y}} $. Therefore, the concentrated, negative log likelihood w.r.t. ${\bf{Y}}$ is written as \cite{DICLAUDIO18}
\begin{equation}
\begin{array}{c}
{\cal L}\left( {{\bf{Y}}|{\left\{ {{{\hat {\bf{E}}}_s},{{\hat {\bf{E}}}_v},{{\hat {\bf{W}}}_s}} \right\}}} \right) = \left( {M - K} \right) \times \\
\left[ {K\ln \left( {\pi {N^{ - 1}}} \right) + \ln \det \left( {{{\bf{Y}}^H}{{\bf{T}}_{ss}}{\bf{Y}}} \right)} \right] + \\
N \times {\mathop{\rm trace}\nolimits} \left[ {{{\bf{Y}}^H}{{\bf{T}}_{vv}}{\bf{Y}}{{\left( {{{\bf{Y}}^H}{{\bf{T}}_{ss}}{\bf{Y}}} \right)}^\dag }} \right]
\end{array}
\label{eqn:marginal_log_likelihood}
\end{equation}
where ${{\bf{T}}_{ss}} = {{{\hat{\bf C}}}^H}{\bf{B}}^H{{\hat {\bf{E}}}_s}\hat {\bf{W}}_s^{ - 2}\hat {\bf{E}}_s^H{\bf{B}}{\hat{\bf C}}$ and ${{\bf{T}}_{vv}} = {{{\hat{\bf C}}}^H}{\bf{B}}^H{{\hat {\bf{E}}}_v}\hat {\bf{E}}_v^H{\bf{B}}{\hat{\bf C}}$.
	
The ML MUSIC solution ${\bf{Y}}$ turns out to be formed by the $K$ generalized eigenvectors ${\bf{y}}_k$ corresponding to the $K$ smallest generalized eigenvalues ${\lambda _k}$ ($k=1,2,\ldots,M$) \cite{DICLAUDIO18}, defined by the Rayleigh quotient 
\begin{equation}
	{\lambda _k} = \frac{{{\bf{y}}_k^H{{\bf{T}}_{vv}}{{\bf{y}}_k}}}{{{\bf{y}}_k^H{{\bf{T}}_{ss}}{{\bf{y}}_k}}} 
\label{eqn:RayleighML}
\end{equation}
after constraining ${\bf{y}}_k$ in the row space of ${\bf{B}}\hat {\bf{C}}$, as in \eqref{eqn:MAP_SVD} and \eqref{eqn:FOCUSS_SVD}.

Within the $O\left( {{N^{ - {1 \mathord{\left/
 {\vphantom {1 2}} \right.
 \kern-\nulldelimiterspace} 2}}}} \right)$ approximation herein adopted for \eqref{eqn:ML_MUSIC_signal} and \eqref{eqn:ML_MUSIC_noise} \cite{STOICA90a}, it is easily verified that ${\lambda _k} = 0$ for $k=1,\ldots,K$ and $Q \ge M$ and that $\bf Y$ by \eqref{eqn:ML_MUSIC_signal} and \eqref{eqn:ML_MUSIC_noise} is the minimum $L_2$ norm solution to the under-determined system
\begin{equation}
	\left[ {\begin{array}{*{20}{c}}
{\hat {\bf{W}}_s^{ - 1}\hat {\bf{E}}_s^H{\bf{B}}\hat {\bf{C}}}\\
{\hat {\bf{E}}_v^H{\bf{B}}\hat {\bf{C}}}
\end{array}} \right]{\bf{Y}} = \left[ {\begin{array}{*{20}{c}}
{{{\bf{I}}_K}}\\
{\bf{0}}
\end{array}} \right] 
	\label{eqn:ML_MUSICsolution}
\end{equation}
which resembles a MAP-SBS iteration \eqref{eqn:MAP_SVD} obtained by a coherent source prior. 

The ML MUSIC solution ${\bf{S}}_{ML} = \hat {\bf{C}}{\bf{Y}}$ has a very high resolution due to the point mass prior \eqref{eqn:CoherentPseudoCovariance}, but generates tight and powerful source \emph{clusters} in the DOA space whose spread reflects the finite sample and quantization uncertainties \cite{LEE12}. However, the presence of spurious DOAs may hamper further interpolation \cite{BUCRIS12} or mislead SBS convergence, because these algorithms require reliable information about source powers. Therefore, a DOA \emph{validation} scheme is required.

\subsection{P-BOOST spectral estimate}
\label{sec:EBOOSTSpectralEstimate}
A robust solution is herein derived directly from \eqref{eqn:SparseSystem} and \eqref{eqn:ML_MUSICsolution} under the FOCUSS diagonal approximation of ${{\bf{R}}_{ss}}$ \eqref{eqn:WeightedFOCUSS}, which is valid at convergence even in coherent scenarios. First the \emph{smoothed source-only pseudo-covariance} ${{\bf{R}}_{ss}}$ is re-defined as
\begin{equation}
{{\bf{R}}_{ss}} = \sum\limits_{q = 1}^Q {{{\bf{D}}_S^2}(q,q){\bf{B}}\left( {:,q} \right)} {\bf{B}}{\left( {:,q} \right)^H}
	\label{eqn:smoothedRss}
\end{equation}
where the current $q$-th source amplitude estimate is
\begin{equation}
{{\bf{D}}_S}(q,q) = {K^{ - {1 \mathord{\left/
 {\vphantom {1 2}} \right.
 \kern-\nulldelimiterspace} 2}}}{\left\| {{{\bf S}_{ML}}\left( {q,:} \right)} \right\|_2} \;.
\label{eqn:Ds}	
\end{equation}

A \emph{leave-one-out cross-validation} \cite{HUBER81} system is drawn from \eqref{eqn:SparseSystem} for $q=1,2,\ldots,Q$ as
\begin{equation}
\begin{array}{c}
{\bf{B}}\left( {:,q} \right){\bf{S}}\left( {q,:} \right) + \left[ {\sum\limits_{j = 1,{\kern 1pt} j \ne q}^Q {{\bf{B}}\left( {:,j} \right)} {{\bf{D}}_S}(j,j){\bf{Y}}\left( {j,:} \right) + {{{\hat{\bf E}}}_v}{\bf{G}}} \right]\\
 = {{{\hat{\bf E}}}_s}{{{\hat{\bf W}}}_s}
\end{array}
\label{eqn:crossvalidation}
\end{equation}
where ${\bf{Y}}\left( {j,:} \right)$ models \emph{interference row vectors} of size $1 \times K$ whose elements are assumed as independent, circular, zero mean and unit variance random variables. Since the entries of $\bf G$ are assumed as i.i.d. circular variables with zero mean and variance $N^{-1}$ as in \eqref{eqn:marginal_log_likelihood}, \eqref{eqn:crossvalidation} is equivalent to a linear fitting with error weighting matrix
\begin{equation}
	{{\bf{C}}_q} = {\left[ {{{\bf{R}}_Q} - {\bf{D}}_S^2\left( {q,q} \right){\bf{B}}\left( {:,q} \right){\bf{B}}{{\left( {:,q} \right)}^H}} \right]^{{{ - 1} \mathord{\left/
 {\vphantom {{ - 1} 2}} \right.
 \kern-\nulldelimiterspace} 2}}}
\label{eqn:errorweighting}
\end{equation}
where
\begin{equation}
	{{\bf{R}}_Q} = {{\bf{R}}_{ss}} + {N^{ - 1}}{{{\hat{\bf E}}}_v}{\hat{\bf E}}_v^H \; .
	\label{eqn:R_Q}
\end{equation}

By multiplying both members of \eqref{eqn:crossvalidation} by ${\bf{R}}_Q^{{{ - 1} \mathord{\left/
 {\vphantom {{ - 1} 2}} \right.
 \kern-\nulldelimiterspace} 2}}$, defining ${{\bf{u}}_q} = {\bf{R}}_Q^{{{ - 1} \mathord{\left/
 {\vphantom {{ - 1} 2}} \right.
 \kern-\nulldelimiterspace} 2}}{\bf{B}}\left( {:,q} \right)$, ${{\bf{U}}_{ \bot q}}$ as the $\left( {M-1} \right)$-th dimensional orthogonal complement to ${{\bf{u}}_q}$ and the partitioned \emph{residual matrix} ${\left[ {\begin{array}{*{20}{c}}
{{\bf t}_q^T}&{{\bf{T}}_{ \bot q}^T}
\end{array}} \right]^T}$ of size $M \times K$, assumed as a random matrix with unit variance elements, straight algebra leads to the linear equation set in Paige form
\begin{equation}
\begin{array}{c}
{{\bf{u}}_q}{\bf{S}}(q,:) + \left[ {\begin{array}{*{20}{c}}
{\displaystyle\frac{{\sqrt {1 - {\bf{D}}_S^2\left( {q,q} \right)\left\| {{{\bf{u}}_q}} \right\|_2^2} }}{{{{\left\| {{{\bf{u}}_q}} \right\|}_2}}}{{\bf{u}}_q}}&{{{\bf{U}}_{ \bot q}}}
\end{array}} \right]\left[ {\begin{array}{*{20}{c}}
{{\bf t}_q}\\
{{{\bf{T}}_{ \bot q}}}
\end{array}} \right]\\
 = {\bf{R}}_Q^{{{ - 1} \mathord{\left/
 {\vphantom {{ - 1} 2}} \right.
 \kern-\nulldelimiterspace} 2}}{\hat {\bf{E}}_s}{\hat {\bf{W}}_s} \;.
\end{array}
\label{eqn:whitened_validation}
\end{equation}

Its generalized LS solution yields \cite{GOLUB89} ${\bf t}_q=0$ and 
\begin{equation}
	{{\bf S}_{LS}}(q,:) = \frac{{{\bf{u}}_q^H{\bf{R}}_Q^{{{ - 1} \mathord{\left/
 {\vphantom {{ - 1} 2}} \right.
 \kern-\nulldelimiterspace} 2}}{{\hat {\bf{E}}}_s}{{\hat {\bf{W}}}_s}}}{{\left\| {{{\bf{u}}_q}} \right\|_2^2}} = \frac{{{\bf{B}}{{\left( {:,q} \right)}^H}{\bf{R}}_Q^{ - 1}{{\hat {\bf{E}}}_s}{{\hat {\bf{W}}}_s}}}{{{\bf{B}}{{\left( {:,q} \right)}^H}{\bf{R}}_Q^{ - 1}{\bf{B}}\left( {:,q} \right)}} \;.
	\label{eqn:EBOOST_estimate}
\end{equation}
which is the BLUE of ${\bf{S}}(q,:)$ for $q=1,2,\ldots,Q$ under the stated assumptions \cite{KAY93}. It is interpreted as the output of a Capon beamformer applied to the weighted signal subspace.

The refined, energy localized \emph{P-BOOST estimate} of the source amplitudes is still given by \eqref{eqn:Ds} as ${{\bf{D}}_S}\left( {q,q} \right) = {K^{ - 1/2}}{\left\| {{{\bf{S}}_{LS}}(q,:)} \right\|_2}$. It has a resolution similar to the Capon MV spectral estimate. The resulting coarse, but high resolution spectral estimator allows a localization of coherent paths and of tight clusters of uncorrelated sources\footnote{In this case, the DOA resolution of MD-MUSIC and of the spectral MUSIC are about the same, since the classical ${{\hat {\bf{E}}}_v}$ herein used converges slowly w.r.t. $N$ \cite{MESTRE08}.}, more than adequate for narrow-band array interpolation on virtual Vandermonde arrays \cite{FRIEDLANDER93,BUCRIS12}, followed by statistically efficient MODE like estimators for DOA super-resolution \cite{STOICA90a}.

For ML or WSF initialization with general array geometries, intra-cluster source super-resolution may be refined by FOCUSS or SBS iterations, quickly converging to a sparse solution, though limited by the codebook DOA quantization. 

\subsection{Remarks}
\label{sec:Remarks}
The smoothed ${{\bf{R}}_{ss}}$ estimate \eqref{eqn:smoothedRss} has numerical rank larger than $K$ and in most cases even of $D$, with a sharp eigenvalue separation if the DOA quantization is smaller than about one tenth of a beamwidth \cite{HUNG88,FRIEDLANDER93,BUCRIS12}. Therefore, some components of ${{\bf{R}}_{ss}}$ leak into ${{{\hat{\bf E}}}_v}$ in the case of coherent sources. However, some of these components originate from $O\left( {{N^{ - {1 \mathord{\left/
 {\vphantom {1 2}} \right.
 \kern-\nulldelimiterspace} 2}}}} \right)$ finite sample and DOA quantization errors and warp the noise subspace of ${{\bf{R}}_{Q}}$ in \eqref{eqn:R_Q}. This phenomenon generates bias and small spurious sources through \eqref{eqn:EBOOST_estimate} \cite{LI05}. Therefore, we developed a cleaning procedure for ${{\bf{R}}_Q}$. 

In particular, a generalized SVD \cite{GOLUB89} yields ${{\bf{R}}_{ss}} = {\bf{F}}{{\bm{\Lambda }}_{ss}}{{\bf{F}}^H}$, ${N^{ - 1}}{{{\hat{\bf E}}}_v}{\hat{\bf E}}_v^H = {\bf{F}}{{\bm{\Lambda }}_{vv}}{{\bf{F}}^H}$ with diagonal, positive semi-definite ${{\bm{\Lambda }}_{ss}}$ and ${{\bm{\Lambda }}_{vv}}$ satisfying ${{\bm{\Lambda }}_{ss}} + {{\bm{\Lambda }}_{vv}} = {{\bf{I}}_M}$. ${\bf{F}}$ is a full rank square matrix of size $M$ \cite{GOLUB89}. In our implementation, we suppress either the generalized components of ${{\bf{R}}_{ss}}$ below the noise level, or the noise components below the signal level. This is accomplished at negligible cost by calculating a diagonal matrix ${{\bm{\Lambda }}_Q}$ with ${{\bm{\Lambda }}_Q}\left( {k,k} \right) = \max \left[ {{{\bm{\Lambda }}_{ss}}\left( {k,k} \right),{{\bm{\Lambda }}_{vv}}\left( {k,k} \right)} \right]$ for $k=1,2,\ldots,M$ and reconstructing
\begin{equation}
	{{\bf{R}}_Q} = {\bf{F}}{{\bf{\Lambda }}_Q}{\bf{F}}^H \; .
	\label{eqn:RQdenoised}
\end{equation}

P-BOOST can be adapted to different signal and noise subspace estimates \cite{JACOVITTI94,STOICA96,GELLI03,MESTRE08,DICLAUDIO18}, if their first-order perturbative model is available. Some of these subspace estimates converge with $N$ \emph{much faster} than the classical ML one \eqref{eqn:CovarianceEVD}, allowing better DOA resolution \cite{MESTRE08} at the price of higher algorithmic complexity.

It is evident that computation of \eqref{eqn:EBOOST_estimate} and subsequent processing can be made on DOAs different from those of the original codebook ${\bf{B}}$. In particular, candidate DOAs can be restricted and finely quantized around the peaks of P-BOOST \cite{HUNG88,DICLAUDIO18}, with benefits for accuracy, resolution and computational load, since source clusters are sharply defined. Moreover, since coherent sources tend to appear in clusters, the OMP search \eqref{eqn:new_codebook_vector} might be limited within a predefined sector around each codebook DOA. Bayesian sector interpolation \cite{BUCRIS12} is straightforward. Finally, \eqref{eqn:EBOOST_estimate} is amenable to be mapped into a rational function for ULAs or harmonically interpolated arrays \cite{RAO89,DORON94A}. However, P-BOOST computation was restricted to codebook DOAs in simulations for fairness. 

\section{Algorithm Summary and Computational Analysis}
\label{sec:ComputationalAnalysis}
The purpose of P-BOOST is two-fold. The first goal is improve coarse DOA estimates for initializing ML or other SBS algorithms under low SNR and coherent scenarios. The second goal is to reduce the \emph{processing latency} with a parallel implementation. Latency is extremely important for coherent super-resolution of source clusters in remote sensing and communications and requires the maximal reduction of iterations and conditional steps, even at the expense of the global number of operations \cite{KUNG85}.
 
In summary, P-BOOST is composed by three fundamental steps:
\begin{description}
	\item [\emph{Step 1}.] \enskip Compute a set of $Q$ independent CSVs by OMP applied to \eqref{eqn:md-music}. 
	\item [\emph{Step 2}.] \enskip Compute \eqref{eqn:ML-MUSIC-mat} and the ML MUSIC solution \eqref{eqn:ML_MUSICsolution}.
	\item [\emph{Step 3}.] \enskip Compute the smoothed covariance \eqref{eqn:RQdenoised} and the P-BOOST estimate \eqref{eqn:EBOOST_estimate}.
\end{description}
Each OMP solution of \eqref{eqn:md-music} in Step 1, of maximum size $M-K-1$, has a bulk cost of $4{Q}{\left( {M - K } \right)^2}$ FLOPs, plus negligible $3{\left( {M - K } \right)^3}$ and $O\left( {M} \right) $ scalar operations for coefficients and likelihood computations for $Q \gg M$. CSVs \eqref{eqn:md-music} can be computed in \emph{parallel} with $O\left( {Q} \right) $ processors after a common online effort of about $MQ^2$ FLOPs for preparing \eqref{eqn:MD-MUSICslice}. In simulations, each OMP task required an average of $2.1$ ms on a $4.2$ GHz PC machine against an average of $7.5$ ms of a numerically stable QRD or SVD based FOCUSS or SBS iteration (rated at a bulk cost of $3M^2Q$ FLOPs \cite{GOLUB89}), due to the low-level vectorization of OMP \cite{CADZOW90,MALLAT93}.

The fitting \eqref{eqn:ML_MUSICsolution} in Step 2 has evidently the same cost of a SBS iteration, plus other $O{\left( {2QM^2} \right)}$ FLOPs for two sparse matrix multiplications to get \eqref{eqn:coherent_solution}. 

Building \eqref{eqn:RQdenoised} in Step 3 has a dominant cost of still $3MQ^2$, plus $QM^2 + O\left( {QMK} \right) $ FLOPs for computing P-BOOST by \eqref{eqn:EBOOST_estimate}. Even this task can be parallelized on $O\left( {M} \right) $ processors.

So the overall latency of P-BOOST can be reduced down to about four equivalent SBS iterations \eqref{eqn:MAP_SVD}. However, the convergence of the FOCUSS-type iterations was uniformly better within the P-BOOST chain, empirically demonstrating its sensitivity to different initializations. FOCUSS converged within about ten iterations up to a relative error of $10^{-7}$ w.r.t. the Frobenius norm of the solution when driven by P-BOOST at any SNR, with a nearly quadratic descent after one or two iterations. A similar number of iterations was required under $10$ dB SNR starting from the $L_2$ norm solution, but with a marginal probability of success. At higher SNR, the same FOCUSS required many more iterations (about $20$ at 20 dB SNR and more than $50$ at $40$ dB SNR) than P-BOOST, with almost linear convergence \cite{LUENBERGER89}.

In conclusion, the latency speedup ratio obtained by a parallel P-BOOST can be of $3-5$ times if followed by FOCUSS. For comparison, the best version of MODE \cite{STOICA90a} required only $5-10$ ms on the same problem and machine, indicating that array interpolation \cite{FRIEDLANDER93} and polynomial rooting \cite{RAO89} remain reference techniques whenever applicable.

\section{AIC Based Source Detection from Spatial Spectra}
\label{sec:AICTypeSourceValidationFromSparseSpatialSpectra}
Because of the inconsistency of quantized DOA estimators, the risk of spurious outcomes is always present and requires a proper source detection stage. Rigorous source number estimation by Information Theoretic Criteria \cite{WAX85} requires the prior computation of \emph{asymptotically efficient} estimates of the source DOAs for the various tested hypotheses \cite{BURNHAM04}. In particular, the DOA quantization bias results in either spurious peaks, or in excess fitting errors and spectral loss of resolution w.r.t. to their ML counterparts. In essence, the subspace fitting error \eqref{eqn:WSF} for different hypotheses significantly decreases for each added candidate source, overcompensating the complexity penalty. This behavior results typically in over-estimation of the source number at medium and high SNRs.

The original formulation of the AIC refers to the \emph{average} negative log-likelihood over a set of observations \cite{AKAIKE74}. The current detection problem has been reformulated in the sparse fitting case, considering only source amplitudes. The key observation is that the selected DOA quantization represents just one possibility within a \emph{dense family of close codebook choices}. Moreover, optimal source validation requires that ${\hat D} < M$ candidate sources are extracted from the codebook after convergence and applied to \eqref{eqn:WSF}, giving origin to a combinatorial, \emph{over-determined} LS fitting \cite{GERSHMAN99}.

The AIC for the $p$-th trial of a single DOA quantization hypothesis ($p=1,2,\ldots,P$), using $K$ signal eigenvectors, depends upon the quantity $MK\log \left( {{{{\mu _p}} \mathord{\left/
 {\vphantom {{{\mu _p}} MK}} \right.
 \kern-\nulldelimiterspace} MK}} \right)$, where ${{\mu _p}}$ is the LS fitting error of \eqref{eqn:WSF}. Now we assume that the WSF is repeated $P$ times under slightly different, independent DOA quantizations and we seek for a \emph{single, compromise fitting coefficient set}. By defining the overall average LS error $\left\langle \mu \right\rangle = {P^{ - 1}}\sum\limits_{p = 1}^P {{\mu _p}} $ and $ {\delta _p} = {\mu _p} - \left\langle \mu \right\rangle $, the average AIC is
\begin{equation}
\frac{1}{P}\sum\limits_{p = 1}^P {MK\log \left( {\frac{{{\mu _p}}}{{MK}}} \right)}  \approx MK\left[ {\log \left( {\frac{{\left\langle \mu  \right\rangle }}{{MK}}} \right) + \frac{1}{P}\sum\limits_{p = 1}^P {\frac{{{\delta _p}}}{{\left\langle \mu  \right\rangle }}} } \right]
\label{eqn:globalAIC}
\end{equation}
up to constant terms and a first order Taylor approximation. If $\frac{1}{P} \sum\limits_{p = 1}^P {\frac{{{\delta _p}}}{{\left\langle \mu \right\rangle }}}$ is negligible or ${\delta _p}$ is a zero mean random variable, the overall WSF solution for source amplitudes can be approximated by concatenating the $P$ LS equation sets.

In the present case, we suppose to find $Q_1<M$ maxima in a generic, discrete spatial spectrum and we search for the source amplitudes of $D \le Q_1$ sources by the overdetermined WSF \cite{CADZOW90}
\begin{equation}
	{{\bf{B}}_D}{{\bf{S}}_D} \approx {{{\hat{\bf E}}}_s}{{{\hat{\bf W}}}_s} 
	\label{eqn:overdeterminedWSFset}
\end{equation}
where the column vectors ${{\bf{B}}_D}\left( {:,d} \right)$ ($d=1,2,\ldots,D$) are extracted from the original codebook $\bf B$.

In general, we have to try all combinations of $D\le Q_1$ peak angles out of the set $\left\{ {{{\bm{\theta }}_1},{{\bm{\theta }}_2}, \ldots ,{{\bm{\theta }}_{{Q_1}}}} \right\}$ \cite{GERSHMAN99}. In a quantized DOA setting, the WSF solution is influenced by the mismatched codebook angles, whose error byproducts tend to mask nearby sources \cite{SWINDLEHURST92,SWINDLEHURST93}. Therefore, for each source combination, we compute a \emph{robust, compromise LS solution} of \eqref{eqn:overdeterminedWSFset} by slightly and randomly moving the selected ${{\bm{\theta }}_d}$ around $\left\{ {{{\bm{\theta }}_1},{{\bm{\theta }}_2}, \ldots ,{{\bm{\theta }}_{{Q_1}}}} \right\}$.

This solution can be analytically approximated by considering that ${{\bf{B}}_D}\left( {:,q} \right) = {\bf{R}}_{vv}^{ - {1 \mathord{\left/
 {\vphantom {1 2}} \right.
 \kern-\nulldelimiterspace} 2}}{\bf{a}}\left( {{{\bm{\theta }}_d}} \right)$ and
\begin{equation}
	{\bf{R}}_{vv}^{ - {1 \mathord{\left/
 {\vphantom {1 2}} \right.
 \kern-\nulldelimiterspace} 2}}{\bf{a}}\left( {{{\bm{\theta }}_d} + {{\bm{\Delta }}_d}} \right) \approx {\bf{R}}_{vv}^{ - {1 \mathord{\left/
 {\vphantom {1 2}} \right.
 \kern-\nulldelimiterspace} 2}}\left[ {{\bf{a}}\left( {{{\bm{\theta }}_d}} \right) + {\nabla _{\bm{\theta }}}\left( {{{\bm{\theta }}_d}} \right){{\bm{\Delta }}_d}} \right]
	\label{eqn:PerturbedSteeringvector}
\end{equation}
in a first order Taylor expansion, where ${\nabla _{\bm{\theta }}}\left( {{{\bm{\theta }}_d}} \right)$ is the row gradient vector of ${\bf{a}}\left( {{{\bm{\theta }}}} \right)$ evaluated at $ {{{\bm{\theta }}}} = {{{\bm{\theta }}_d}}$.

Assuming that each codebook DOA perturbation ${\bm{\Delta }}_d$ for $d=1,2,\ldots,D$ is a random multivariate vector with independent components, uniformly distributed within the DOA quantization interval, the average solution of \eqref{eqn:overdeterminedWSFset} is found from the \emph{diagonally} regularized LS equation set
\begin{equation}
	\left[ {\begin{array}{*{20}{c}}
{{{\bf{B}}_D}}\\
{{{\bf{G}}_{\bm{\theta } }}}
\end{array}} \right]{{\bf{S}}_D} \approx \left[ {\begin{array}{*{20}{c}}
{{{{\hat{\bf E}}}_s}{{{\hat{\bf W}}}_s}}\\
{\bf{0}}
\end{array}} \right]
\label{eqn:RegularizedWSFset}
\end{equation}
where ${\bf{G}}_{\bm{\theta } }\left( {d,d} \right)^2 = E\left\{ {{\bm{\Delta }}_d^T{\nabla _{\bm{\theta }}}{{\left( {{{\bm{\theta }}_d}} \right)}^H}{\bf{R}}_{vv}^{ - 1}{\nabla _{\bm{\theta }}}\left( {{{\bm{\theta }}_d}} \right){{\bm{\Delta }}_d}} \right\}$.

The LS fitting error of \eqref{eqn:RegularizedWSFset} estimates the $\left\langle \mu \right\rangle $ for the approximate AIC computation by \eqref{eqn:RegularizedWSFset}
\begin{equation}
\mathrm{AIC}\left( D \right) = 2\left\{ {MK\log \left( {\frac{{\pi \left\langle \mu  \right\rangle }}{{MK}}} \right) + MK + \left( {2DK + 1} \right)} \right\}\; .
\label{eqn:GeneralizationAIC}	
\end{equation}

The generalized error variance ${M^{ - 1}}{K^{ - 1}}\left\langle \mu \right\rangle$ was found to dominate both the higher order AIC components \cite{BURNHAM04} and the WSF errors before regularization \eqref{eqn:RegularizedWSFset} for typical DOA quantization steps. In particular, the regularization forces toward zero the signal components comparable to the predicted level of DOA mismatch error. In our implementation, peaks were sorted in a non-increasing magnitude to get a worst case estimate of the mismatch level. $\mathrm{AIC}\left( D \right)$ was sequentially evaluated with fitting orders ranging from $0$ to $Q_1$. In simulations, the peak angles and magnitudes were refined through a classical three-point parabolic interpolation \cite{JACOVITTI93} of the logarithm of the spatial spectrum. Each estimate was reassigned to the closest codebook DOA.

\section{Statistical validation of sparse solutions}
\label{sec:StatisticalValidationOfSparseSolutions}
The validation of DOA quantized estimators \cite{WIPF07} poses several issues, especially by considering that their computational requirements grow at least proportionally to $Q$ and that bias and variance figures of classical parametric DOA estimators at typical SNR are often smaller by orders of magnitude w.r.t. practical angle quantization. 

On the other side, by fixing the codebook angles and exploiting prior Bayesian information, these methods can provide extremely robust DOA estimates at low SNR, when classical parametric estimators \cite{STOICA90a,VIBERG91} diverge from the CRB.

Even at very low SNR, a fraction of DOA estimates tends to cluster around the true values. The remaining estimates are spread through the field of view, represent \emph{gross errors} or \emph{spurious sources} and their frequency decreases with the SNR. This observation leads to model sample DOA estimates by a \emph{contaminated distribution} \cite{HUBER81}. DOA estimates within the WSF or ML trust region lead to safe convergence, occur with probability $P_d$ and their conditional distribution approaches normality if the estimation variance and the quantization step are much smaller than the trust region itself. Gross errors, herein referred to as \emph{failures}, occur with probability $1-P_d$ and have an unknown, high variance distribution. In tracking applications, $P_d$ is easily related to the number of attempts to obtain a detection.

In the simulation analysis reported in Sect. \ref{sec:ComputerSimulations}, the candidate DOA estimates were first \emph{paired} to the true ones by an extended, Root Mean Squared Error (RMSE) based \emph{Hungarian algorithm} \cite{MUNKRES57}. We decided to mark as \emph{failures} either \emph{missing} DOA estimates (i.e., paths not paired to any DOA estimate), or the presence of paired estimates, whose absolute DOA error exceeded a given threshold, determined after some WSF trials. 

Overestimation phenomena of the source number (i.e., the presence of candidate DOAs not assigned to any path) were not penalized, since WSF estimators can deal with this issue \cite{VIBERG91,GERSHMAN99}. Nevertheless, they can lead to the pairing of DOAs that might be discarded by a statistical criterion. 

The second parameter of interest is the spread of the paired candidate DOAs around the true values. Sparse solvers have bias but may obtain a zero DOA variance over a certain SNR threshold \cite{GORODNITSKY97}. The \emph{trimmed, conditional} RMSE, computed on the subset of successfully paired DOA estimates, was adopted in this work. However, RMSE is far from ideal, since the trust region of WSF is almost symmetric and penalizes biased coarse estimates. Moreover, this conditional RMSE can be smaller than the square root of the (unbiased) CRB at low SNR, because of the \emph{trimming} of the sample distribution \cite{HUBER81} and because sparse solvers may exploit prior information and are generally biased.

\section{Computer simulations}
\label{sec:ComputerSimulations}
After preliminary trials, we chose to compare the MAP/SBS version of FOCUSS \eqref{eqn:MAP_SVD} \cite{WIPF07}, boosted by the minimum norm $L_2$ solution, a similar SBS with off-grid DOA refinement \cite{YANG13}, the OMP with known path number $D$ \cite{CADZOW90}, the P-BOOST \eqref{eqn:EBOOST_estimate} alone, the P-BOOST chained with MAP FOCUSS and eventually its WSF AIC validation. The $L_1$ penalized sparse solution was also tried, but it was unable to reliably resolve DOAs in most scenarios and exhibited a very slow numerical convergence.

Two versions of MODE, the original one \cite{STOICA90a} and an improved variant with centro-symmetric internal equations and polynomial vector\footnote{It is roughly equivalent to a forward-backward covariance smoothing \cite{GERSHMAN99}, but it is statistically efficient even after noise whitening and gave similar performance of local WSF descent \cite{CADZOW90,VIBERG91}.} were used as benchmark, programmed for the exact $D$. MODE requires Vandermonde arrays and has been chosen for its analytical, globally converging WSF formulation, but it furnishes a reference for \emph{perfect} array interpolation onto a virtual ULA \cite{FRIEDLANDER93}. However, accurate interpolation can be made only within small angular sectors and with prior knowledge of the source DOAs and powers \cite{BUCRIS12}. Therefore, a booster like OMP \cite{CADZOW90} or P-BOOST is always required. 

A narrow-band $M=8$ sensor ULA, inter-spaced by $0.5$ wavelengths, was simulated by choosing $180$ DOA angles for the codebook $\bf B$, ranging from $-89.5 ^\circ$ to $89.5 ^\circ$ in steps of one degree, referred to broadside, exceeding by about four times established array sampling criteria \cite{DORON94A,BUCRIS12}. The quantization DOA RMSE floor was $0.29^\circ$. Additive sensor noise was spherical, zero mean, circular Gaussian distributed. The SNR was referred to each source and sensor. Source DOAs were chosen midway between codebook DOAs in order to maximize quantization errors.

Two environments were simulated, both with equi-powered, circular Gaussian sources impinging from $8 ^\circ$, $13 ^\circ$, $33 ^\circ$ and $37 ^\circ$ \cite{HUNG88,DICLAUDIO18}. In the first setting, sources were uncorrelated, while in the second the arrivals at $13 ^\circ$ and $37 ^\circ$ were fully coherent with those coming from $8 ^\circ$ and $33 ^\circ$, as in \cite{DICLAUDIO18}. One thousand trials were run for each SNR, processing $N=100$ independent snapshots. The SCM rank was fixed to four in the uncorrelated case and to two in the coherent case, to avoid unwanted conditioning of path number estimates. The failure threshold was set at seven degrees.

\begin{figure}[tb]
	\centering
		\includegraphics[width=3.2 in]{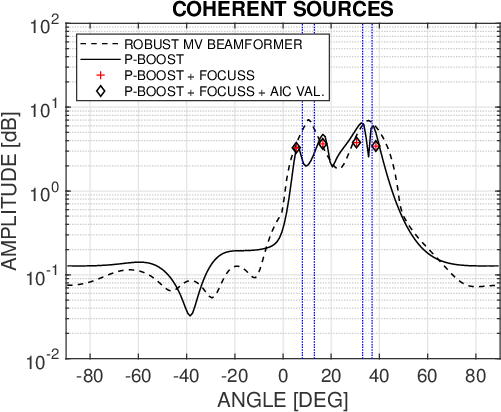}
	\caption {The pseudospectra of the various P-BOOST stages compared with the output of a robust MV beamformer under coherent arrivals at $\rm{SNR} = 10$ dB. Vertical dotted lines mark the true DOAs.}
	\label{fig:Fig1}
\end{figure}

Fig. \ref{fig:Fig1} shows sample steps of the P-BOOST plus FOCUSS chain in the coherent case at $10$ dB SNR, compared with the sample spatial spectrum of a robust distortion-less MV beamformer \cite{GRIFFITHS82,LI05}. The beamformer used covariance shrinkage \cite{LEDOIT04}, a linear zero gradient constraint and constrained the $L_2$ norm of the weight vector below $\sqrt{2}$ for contrasting, respectively, finite sample errors, DOA mismatch and signal coherence \cite{GRIFFITHS82,LI05}.

The robust beamformer exhibited strong losses of DOA resolution, but delineated fairly well the two source clusters, useful for building interpolation matrices \cite{FRIEDLANDER93,BUCRIS12}. In this case, the P-BOOST pseudo-spectrum \eqref{eqn:EBOOST_estimate} resolved the four sources, while subsequent FOCUSS refinement and the final AIC validation isolated the peaks with a certain bias w.r.t. the actual DOAs, marked by dotted, vertical lines in Fig. \ref{fig:Fig1}.

\begin{figure}[tb]
	\centering
		\includegraphics[width=3.2 in]{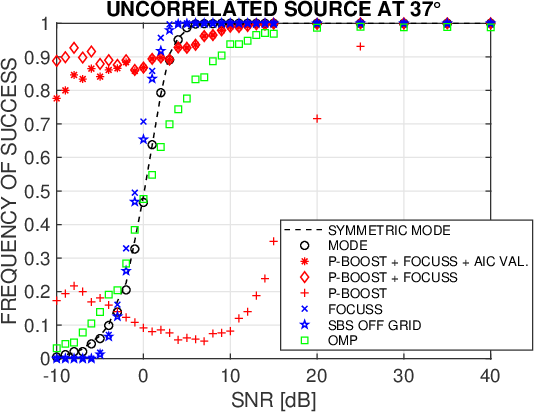}
	\caption {Sample rates of overall success of the various DOA boosters w.r.t. the SNR in the uncorrelated case.}
	\label{fig:Fig2}
	\vspace{12pt}
		\centering
		\includegraphics[width=3.2 in]{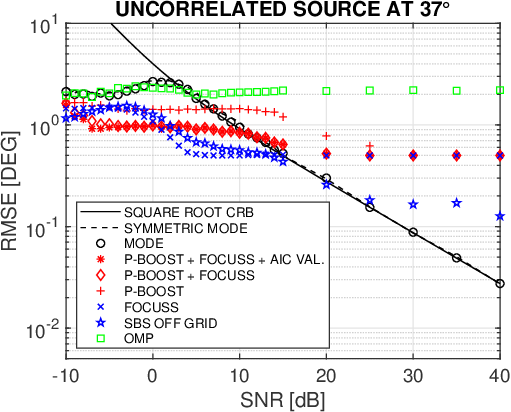}
	\caption {Conditional sample DOA RMSE of the various estimators for the uncorrelated source located at $37 ^\circ$.}
	\label{fig:Fig3}
\end{figure}

\begin{figure}[tb]
	\centering
		\includegraphics[width=3.2 in]{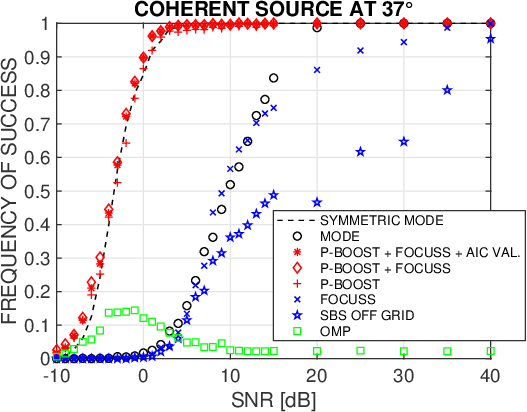}
	\caption {Sample rates of overall success of the various DOA boosters w.r.t. the SNR in the coherent case.}
	\label{fig:Fig4}
	\vspace{12pt}
		\centering
		\includegraphics[width=3.2 in]{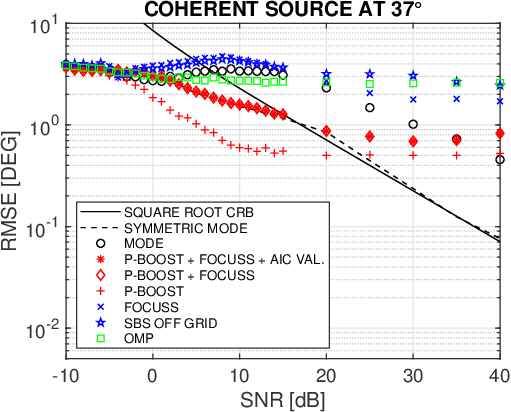}
	\caption {Conditional sample DOA RMSE of the various estimators for the coherent source located at $37 ^\circ$.}
	\label{fig:Fig5}
\end{figure}

The \emph{overall} rate of success $P_d$ of the DOA boosting and the conditional RMSE of compared estimators w.r.t. the SNR for the most difficult source located at $37 ^\circ$ were depicted in Figs. \ref{fig:Fig2} and \ref{fig:Fig3} for the uncorrelated case and in Figs. \ref{fig:Fig4} and \ref{fig:Fig5} for the coherent case.

\begin{figure}[tb]
	\centering
		\includegraphics[width=3.2 in]{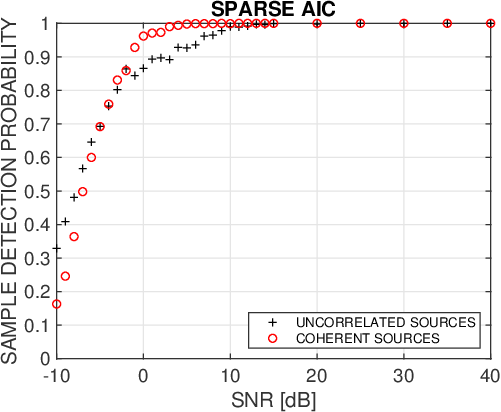}
	\caption {Sample detection probability curves versus SNR for the final AIC detection stage of P-BOOST for the uncorrelated and the coherent settings.}
	\label{fig:Fig6}
\end{figure}

The curves of the success rate $P_d$ of correct path detection ($D=4$) versus SNR of the final AIC detection stage of P-BOOST are reported in Fig. \ref{fig:Fig6} for both uncorrelated and coherent settings. By comparison with the symmetric MODE performance, the $P_d$ of the P-BOOST chain is adequate for ML or WSF initialization in the useful SNR range.

The symmetric MODE confirmed its statistical efficiency and remains the reference DOA estimator Vandermonde arrays. The classical MODE \cite{STOICA90a} exhibited the known local convergence weakness \cite{DICLAUDIO18} in the coherent case, due to the appearance of two small and close singular values in its internal system matrix. The SBS with off grid DOA refinement departed from  MODE at high SNR, exhibited a slow and uncertain convergence at any SNR ($50$-$1000$ iterations) and failed in the coherent case. The problem resembled that of OMP \cite{CADZOW90} and was essentially due to a strong spurious peak at about $20 ^\circ$ in the coherent case, capable to hamper the local WSF convergence. These estimators do not look competitive w.r.t. the other solutions, that use a coarse DOA estimate before a local WSF optimization. The OMP obtained a marginal coherent resolution only between $-5$ and $0$ dB SNR, where the classical beamforming was nearly optimal \cite{LI05}. The kernel superposition issue was instead relevant at higher SNRs.

For other array geometries, WSF needs a booster and the P-BOOST chain is clearly the best choice among DOA quantized estimators, that exhibited the typical RMSE \emph{plateau} at high SNR. P-BOOST alone exhibited loss of resolution of the ($33 ^\circ,37 ^\circ$) source pair below $20$ dB SNR in the uncorrelated case, due in part to the path coherence assumption (i.e., an \emph{over-parametrized} model), which required some FOCUSS refinements, and in part to the P-BOOST spectrum computation only at codebook DOAs. This performance nearly coincided with that of spectral MUSIC \cite{SCHMIDT86}, not reported in the graph. For the same reason, P-BOOST excelled in the coherent case. 

Interestingly, the AIC validation stage essentially did not cause performance losses to the P-BOOST chain with respect to the oracle-driven selection. The typical AIC over-fitting at high SNR \cite{WAX85} was never observed after P-BOOST (see also Figs. \ref{fig:Fig3} and \ref{fig:Fig5}), due to the cited quantization mismatch dominance and to the scarcity of spurious peaks. At very low SNR, most spurious peaks were suppressed together with some valid DOA estimates. However, the empirical detection curves of the P-BOOST chains were influenced by the cited appearance of spurious close sources producing legitimate DOA matches at very low SNR. In particular, double peaks tended to coalesce at higher SNR before a stable resolution.

In addition to the previous fixed DOA examples, we assessed the estimator behavior in a randomly changing environment. Separate source clusters carried un-correlated signals and were generated according to a marked Matern process \cite{BADDELEY07}, adapted for far field simulation. The number of clusters followed a Poisson law of intensity one and the cluster DOA centers were uniformly distributed between $-60 ^\circ$ and $60^\circ$\footnote{Extreme angles in a ULA have diverging CRB for DOA, impairing source identifiability \cite{KAY93}. These angles are typically excluded by directional sensors and sector interpolation \cite{FRIEDLANDER93}.}.  

Within each cluster, random sources were uniformly located within a circular disk around the center, that produced a DOA cluster sector of $18 ^\circ$ width. The number of intra-cluster coherent sources followed a Poisson distribution of intensity two. Sources amplitudes were independent and Rayleigh distributed with unit average power. The background spherical noise variance was adjusted to give the specified SNR at each sensor w.r.t. a unit power source. In each cluster, the source correlation coefficients were calculated according to the propagation laws, assuming a sinc-type temporal signal correlation with a correlation radius at the first zero equal to $1.5$ times the disk radius. $20,000$ independent trials were run for each test SNR. The signal subspace rank was estimated by the enhanced AIC of \cite{NADAKUDITI08}. Missed signal detections by AIC (i.e., a null estimated signal subspace rank) and effects of array overloading (i.e., $D \ge M$) were included in the failure count.

Results are displayed in Table \ref{table:RandomEnvironment} in terms of $P_d$ and of the conditional RMSE of the DOAs successfully paired within the tolerance of seven degrees, deemed adequate for WSF local convergence. The small conditional RMSE of $L_1$ at low SNR is a byproduct of its low $P_d$ and does not indicate a better overall precision. Therefore, in Table \ref{table:RandomEnvironment} we reported the conditional RMSE computed on the fraction of the smallest DOA errors, corresponding to the worst $P_d$ measured across tested estimators. 

In this environment, somewhat less demanding than previous ones for average source spacing and correlation, results were intermediate between those of uncorrelated and coherent tests. The two MODE estimators and the OMP knew the true source number and obtained high $P_d$, but scored a rather high conditional RMSE. Among the sparse solvers, P-BOOST alone reached the best $P_d$ and overall RMSE at very low SNR as in the previous coherent case. At higher SNR, the resolution limits of the quantized Capon-like P-BOOST estimate \eqref{eqn:EBOOST_estimate} dominated the RMSE. Surprisingly, FOCUSS initialized by P-BOOST gave slightly worse results at low SNR because of the larger bias and the suppression of some valid sources. 

SBS estimators starting from the $L_2$ solution were weak below $5$ dB SNR and obtained similar performance. However, between $5$ and $10$ dB SNR, the off grid SBS was able to reach better DOA RMSE even than MODE (i.e., WSF), as in the previous uncorrelated case, due to the bounds imposed on the DOA parameter set. 

The AIC detection stage is a valid shortcut to bypass the full WSF hypothesis testing \cite{GERSHMAN99}, since it introduced some detection losses only around $-5$ dB SNR. 

\begin{table*}[tb]
\centering
\renewcommand{\arraystretch}{1.}
\setlength\tabcolsep{6pt}
\caption{Sample detection probability and conditional DOA RMSE for the random environment test}
\label{table:RandomEnvironment}
	\centering
		\begin{tabular}{|c|c|c|c|c|c|c|c|c|c|c|c|c|}
			\hline
			{SNR} & \multicolumn{3}{c|}{$-5$ dB} & \multicolumn{3}{c|}{$0$ dB} & \multicolumn{3}{c|}{$5$ dB} & \multicolumn{3}{c|}{$10$ dB} \tabularnewline 
			\hline
			{DOA Estimator} & $P_d$ & RMSE & RMSE & $P_d$ & RMSE & RMSE & $P_d$ & RMSE & RMSE & $P_d$ & RMSE & RMSE \tabularnewline 
				& & & $11\%$ & & & $20\%$ & & & $30\%$ & & & $39\%$ \tabularnewline
			\hline
			{MODE} & $0.26$ & $2.29^\circ$ & $0.49^\circ$ & $0.41$ & $2.00^\circ$ & $0.47^\circ$ & $0.51$ & $1.69^\circ$ & $0.46^\circ$ & $0.58$ & $1.46^\circ$ & $0.45^\circ$ \tabularnewline
			{Symmetric MODE} & $0.28$ & $2.40^\circ$ & $0.49^\circ$ & $0.44$ & $2.16^\circ$ & $0.45^\circ$ & $0.54$ & $1.90^\circ$ & $0.45^\circ$ & $0.61$ & $1.71^\circ$ & $0.45^\circ$ \tabularnewline
			{OMP} & $0.31$ & $2.37^\circ$ & $0.42^\circ$ & $0.47 $ & $2.22^\circ$ & $0.44^\circ$ & $0.59$ & $2.14^\circ$ & $0.51^\circ$ & $\bf {0.68}$ & $2.12^\circ$ & $0.61^\circ$ \tabularnewline
			\hline
			{$L_1$ sparse solution} & $0.11$ & $1.44^\circ$ & $1.44^\circ$ & $0.20$ & $\bf {1.30}^\circ$ & $1.30^\circ$ & $0.30$ & $\bf {1.10}^\circ$ & $1.10^\circ$ & $0.39$ & $\bf {1.03}^\circ$ & $1.03^\circ$ \tabularnewline
			{$L_2$ solution + FOCUSS} & $0.16$ & ${\bf 1.36}^\circ$ & $0.64^\circ$ & $0.33$ & $1.41^\circ$ & $0.49^\circ$ & $0.48$ & $1.37^\circ$ & $0.46^\circ$ & $0.57$ & $1.29^\circ$ & $0.47^\circ$ \tabularnewline
			{$L_2$ solution + SBS off grid} & $0.20$ & $1.56^\circ$ & $0.53^\circ$ & $0.34$ & $1.43^\circ$ & $0.47^\circ$ & $0.48$ & $1.36^\circ$ & $\bf {0.43}^\circ$ & $0.57$ & $1.26^\circ$ & $\bf{0.43}^\circ$ \tabularnewline
			\hline	
			{P-BOOST} & $\bf{0.32}$ & $2.16^\circ$ & $\bf{0.39}^\circ$ & $\bf {0.51}$ & $2.11^\circ$ & $\bf{0.42}^\circ$ & $\bf {0.60}$ & $2.02^\circ$ & $0.51^\circ$ & $0.63$ & $1.91^\circ$ & $0.58^\circ$ \\
			{P-BOOST + FOCUSS} & $0.30 $ & $1.91^\circ$ & $0.40^\circ$ & $0.44$ & $1.77^\circ$ & $0.43^\circ$ & $0.49$ & $1.52^\circ$ & $0.54^\circ$ & $0.52$ & $1.35^\circ$ & $0.61^\circ$ \tabularnewline
			{P-BOOST + FOCUSS + AIC} & $0.27$ & $1.90^\circ$ & $0.46^\circ$ & $0.43$ & $1.78^\circ$ & $0.45^\circ$ & $0.49$ & $1.54^\circ$ & $0.54^\circ$ & $0.52$ & $1.35^\circ$ & $0.61^\circ$ \tabularnewline
			\hline			
		\end{tabular}
\end{table*}

\section{Conclusion}
\label{sec:Conclusion}
The proposed P-BOOST scheme allows to initialize narrow-band WSF problems with arbitrary arrays. It scores high performance, even at low SNR and in the presence of multiple coherent paths, and offers a high degree of parallelism for low latency applications. P-BOOST proves that the estimates of existing sparse solvers and WSF itself can be improved by a proper initialization, based on the array signal model and revamped classical paradigms. In addition, P-BOOST based array interpolation on tight DOA clusters, followed by MODE, is a nearly asymptotically efficient DOA estimator. Future works will be directed to investigate alternative implementations of P-BOOST, as well as other detection schemes.
\ifCLASSOPTIONcaptionsoff
 \newpage
\fi

%\IEEEtriggeratref{13}


\bibliographystyle{IEEEtran}
\bibliography{IEEEabrv,imageprocessing,arrayprocessing}

\begin{thebibliography}{10}
\providecommand{\url}[1]{#1}
\csname url@samestyle\endcsname
\providecommand{\newblock}{\relax}
\providecommand{\bibinfo}[2]{#2}
\providecommand{\BIBentrySTDinterwordspacing}{\spaceskip=0pt\relax}
\providecommand{\BIBentryALTinterwordstretchfactor}{4}
\providecommand{\BIBentryALTinterwordspacing}{\spaceskip=\fontdimen2\font plus
\BIBentryALTinterwordstretchfactor\fontdimen3\font minus
  \fontdimen4\font\relax}
\providecommand{\BIBforeignlanguage}[2]{{%
\expandafter\ifx\csname l@#1\endcsname\relax
\typeout{** WARNING: IEEEtran.bst: No hyphenation pattern has been}%
\typeout{** loaded for the language `#1'. Using the pattern for}%
\typeout{** the default language instead.}%
\else
\language=\csname l@#1\endcsname
\fi
#2}}
\providecommand{\BIBdecl}{\relax}
\BIBdecl

\bibitem{VANTREES02}
H.~V. Trees, Ed., \emph{Detection, Estimation, and Modulation Theory, Part IV,
  Optimum Array Processing}, 1st~ed.\hskip 1em plus 0.5em minus 0.4em\relax New
  York: John Wiley and Sons, Apr. 2002.

\bibitem{BOHME85}
J.~F. Bohme, ``Source-parameter estimation by approximate maximum likelihood
  and nonlinear regression,'' \emph{IEEE Journal of Oceanic Engineering},
  vol.~10, no.~3, pp. 206--212, July 1985.

\bibitem{STOICA90a}
P.~Stoica and K.~C. Sharman, ``Maximum likelihood methods for
  direction-of-arrival estimation,'' \emph{IEEE Transactions on Acoustics,
  Speech, and Signal Processing}, vol.~38, no.~7, pp. 1132--1143, Jul. 1990.

\bibitem{VIBERG91}
M.~Viberg, B.~Ottersten, and T.~Kailath, ``Detection and estimation in sensor
  arrays using weighted subspace fitting,'' \emph{IEEE Transactions on Signal
  Processing}, vol.~39, no.~11, pp. 2436 --2449, Nov. 1991.

\bibitem{DEMOOR93}
B.~{De Moor}, ``The singular value decomposition and long and short spaces of
  noisy matrices,'' \emph{IEEE Transactions on Signal Processing}, vol.~41,
  no.~9, pp. 2826--2838, Sept. 1993.

\bibitem{SCHMIDT86}
R.~Schmidt, ``Multiple emitter location and signal parameter estimation,''
  \emph{IEEE Transactions on Antennas and Propagation}, vol.~34, no.~3, pp.
  276--280, March 1986.

\bibitem{WAX94}
M.~Wax and J.~Sheinvald, ``Direction finding of coherent signals via spatial
  smoothing for uniform circular arrays,'' \emph{IEEE Transactions on Antennas
  and Propagation}, vol.~42, no.~5, pp. 613--620, May 1994.

\bibitem{FRIEDLANDER93}
B.~Friedlander and A.~Weiss, ``Direction finding for wide-band signals using an
  interpolated array,'' \emph{IEEE Transactions Signal Processing}, vol.~41,
  no.~4, pp. 1618--1635, April 1993.

\bibitem{WEISS95}
A.~J. Weiss, B.~Friedlander, and P.~Stoica, ``Direction-of-arrival estimation
  using {MODE} with interpolated arrays,'' \emph{IEEE Transactions on Signal
  Processing}, vol.~43, no.~1, pp. 296--300, Jan 1995.

\bibitem{LUENBERGER89}
D.~Luenberger, \emph{Linear and nonlinear programming}, 2nd~ed.\hskip 1em plus
  0.5em minus 0.4em\relax Addison Wesley, 1989.

\bibitem{CADZOW90}
J.~Cadzow, ``Multiple source location - {T}he signal subspace approach,''
  \emph{IEEE Transactions on Acoustics Speech and Signal Processing}, vol.~38,
  no.~7, pp. 1110--1125, July 1990.

\bibitem{WAX85}
M.~Wax and T.~Kailath, ``Detection of signals by {I}nformation {T}heoretic
  {C}riteria,'' \emph{IEEE Transactions on Acoustics Speech and Signal
  Processing}, vol.~33, no.~2, pp. 387--392, April 1985.

\bibitem{NADAKUDITI08}
R.~R. Nadakuditi and A.~Edelman, ``Sample eigenvalue based detection of
  high-dimensional signals in white noise using relatively few samples,''
  \emph{IEEE Transactions on Signal Processing}, vol.~56, no.~7, pp.
  2625--2638, July 2008.

\bibitem{LI05}
J.~Li and P.~Stoica, Eds., \emph{Robust adaptive beamforming}, ser. Wiley
  Series in Telecommunications and Signal Processing.\hskip 1em plus 0.5em
  minus 0.4em\relax Hoboken, NJ, USA: J. Wiley and Sons, Oct. 2005.

\bibitem{ROY89}
R.~Roy and K.~Kailath, ``{ESPRIT} - estimation of signal parameter via
  rotational invariance techniques,'' \emph{IEEE Transactions on Acoustics
  Speech and Signal Processing}, vol.~37, no.~7, pp. 984--995, July 1989.

\bibitem{MASSA15}
A.~Massa, P.~Rocca, and G.~Oliveri, ``Compressive sensing in electromagnetics -
  a review,'' \emph{IEEE Antennas and Propagation Magazine}, vol.~57, no.~1,
  pp. 224--238, Feb 2015.

\bibitem{GORODNITSKY97}
I.~F. Gorodnitsky and B.~D. Rao, ``Sparse signal reconstruction from limited
  data using {FOCUSS}: a re-weighted minimum norm algorithm,'' \emph{IEEE
  Transactions on Signal Processing}, vol.~45, no.~3, pp. 600--616, Mar 1997.

\bibitem{MALIOUTOV05}
D.~Malioutov, M.~Cetin, and A.~S. Willsky, ``A sparse signal reconstruction
  perspective for source localization with sensor arrays,'' \emph{IEEE
  Transactions on Signal Processing}, vol.~53, no.~8, pp. 3010--3022, Aug 2005.

\bibitem{MALLAT93}
S.~G. Mallat and Z.~Zhang, ``Matching pursuits with time-frequency
  dictionaries,'' \emph{IEEE Transactions on Signal Processing}, vol.~41,
  no.~12, pp. 3397--3415, Dec 1993.

\bibitem{GERSHMAN99}
\BIBentryALTinterwordspacing
A.~B. Gershman and P.~Stoica, ``New {MODE}-based techniques for direction
  finding with an improved threshold performance,'' \emph{Signal Processing},
  vol.~76, no.~3, pp. 221 -- 235, 1999. [Online]. Available:
  \url{http://www.sciencedirect.com/science/article/pii/S0165168499000110}
\BIBentrySTDinterwordspacing

\bibitem{WIPF07}
D.~P. Wipf and B.~D. Rao, ``An empirical {B}ayesian strategy for solving the
  simultaneous sparse approximation problem,'' \emph{IEEE Transactions on
  Signal Processing}, vol.~55, no.~7, pp. 3704--3716, July 2007.

\bibitem{FANNJIANG11}
\BIBentryALTinterwordspacing
A.~C. Fannjiang, ``The {MUSIC} algorithm for sparse objects: a compressed
  sensing analysis,'' \emph{Inverse Problems}, vol.~27, no.~3, p. 035013, 2011.
  [Online]. Available: \url{http://stacks.iop.org/0266-5611/27/i=3/a=035013}
\BIBentrySTDinterwordspacing

\bibitem{LEE12}
K.~{Lee}, Y.~{Bresler}, and M.~{Junge}, ``Subspace methods for joint sparse
  recovery,'' \emph{IEEE Transactions on Information Theory}, vol.~58, no.~6,
  pp. 3613--3641, June 2012.

\bibitem{YANG13}
Z.~{Yang}, L.~{Xie}, and C.~{Zhang}, ``Off-grid direction of arrival estimation
  using sparse b{}ayesian inference,'' \emph{IEEE Transactions on Signal
  Processing}, vol.~61, no.~1, pp. 38--43, Jan 2013.

\bibitem{AKAIKE74}
H.~Akaike, ``A new look at the statistical model identification,''
  \emph{Automatic Control, IEEE Transactions on}, vol.~19, no.~6, pp. 716--723,
  Dec 1974.

\bibitem{BURNHAM04}
\BIBentryALTinterwordspacing
K.~P. Burnham and D.~R. Anderson, ``Multimodel inference: Understanding {AIC}
  and {BIC} in model selection,'' \emph{Sociological Methods and Research},
  vol.~33, no.~2, pp. 261--304, 2004. [Online]. Available:
  \url{http://smr.sagepub.com/content/33/2/261.abstract}
\BIBentrySTDinterwordspacing

\bibitem{BUCRIS12}
Y.~Bucris, I.~Cohen, and M.~A. Doron, ``Bayesian focusing for coherent wideband
  beamforming,'' \emph{IEEE Transactions on Audio, Speech, and Language
  Processing}, vol.~20, no.~4, pp. 1282--1296, May 2012.

\bibitem{KUNG85}
S.~{Kung}, ``Vlsi array processors,'' \emph{IEEE ASSP Magazine}, vol.~2, no.~3,
  pp. 4--22, Jul 1985.

\bibitem{DICLAUDIO18}
E.~D.~D. Claudio, R.~Parisi, and G.~Jacovitti, ``Space time {MUSIC}: Consistent
  signal subspace estimation for wideband sensor arrays,'' \emph{IEEE
  Transactions on Signal Processing}, vol.~66, no.~10, pp. 2685--2699, May
  2018.

\bibitem{HARMANCI00}
K.~Harmanci, J.~Tabrikian, and J.~Krolik, ``Relationships between adaptive
  minimum variance beamforming and optimal source localization,'' \emph{IEEE
  Transactions on Signal Processing}, vol.~48, no.~1, pp. 1--12, Jan. 2000.

\bibitem{HUNG88}
H.~Hung and M.~Kaveh, ``Focussing matrices for coherent signal-subspace
  processing,'' \emph{IEEE Transactions on Acoustics Speech and Signal
  Processing}, vol.~36, no.~8, pp. 1272--1281, Aug. 1988.

\bibitem{DICLAUDIO05}
E.~D. {Di Claudio}, ``Asymptotically perfect wideband focusing of multi-ring
  circular arrays,'' \emph{IEEE Transactions on Signal Processing}, vol. 53,
  Part 1, no.~10, pp. 3661--3673, Oct. 2005.

\bibitem{RAO89}
B.~D. Rao and K.~Hari, ``Performance analysis of root-{MUSIC},'' \emph{IEEE
  Transactions on Acoustics Speech and Signal Processing}, vol.~37, no.~12, pp.
  1939--1949, Dec. 1989.

\bibitem{GOLUB89}
G.~Golub and C.~V. Loan, \emph{Matrix Computations}, 2nd~ed.\hskip 1em plus
  0.5em minus 0.4em\relax Baltimore, USA: John Hopkins University Press, 1989.

\bibitem{HOFFBECK96}
J.~P. Hoffbeck and D.~A. Landgrebe, ``Covariance matrix estimation and
  classification with limited training data,'' \emph{IEEE Transactions on
  Pattern Analysis and Machine Intelligence}, vol.~18, no.~7, pp. 763--767,
  July 1996.

\bibitem{LEDOIT04}
O.~Ledoit and M.~Wolf, ``A well-conditioned estimator for large-dimensional
  covariance matrices,'' \emph{Journal of Multivariate Analysis}, vol.~88,
  no.~2, pp. 365--411, 2004.

\bibitem{MESTRE08}
X.~Mestre and M.~A. Lagunas, ``Modified subspace algorithms for {DoA}
  estimation with large arrays,'' \emph{IEEE Transactions on Signal
  Processing}, vol.~56, no.~2, pp. 598--614, Feb 2008.

\bibitem{DORON93}
M.~Doron, A.~J. Weiss, and H.~Messer, ``Maximum-likelihood direction finding of
  wide-band sources,'' \emph{IEEE Transactions on Signal Processing}, vol.~41,
  no.~1, pp. 411--414, Jan. 1993.

\bibitem{KAY93}
S.~M. Kay, \emph{Fundamentals of Statistical Signal Processing: Estimation
  Theory}.\hskip 1em plus 0.5em minus 0.4em\relax Upper Saddle River, NJ, USA:
  Prentice-Hall, Inc., 1993.

\bibitem{DONOHO03}
\BIBentryALTinterwordspacing
D.~L. Donoho and M.~Elad, ``Optimally sparse representation in general
  (nonorthogonal) dictionaries via {L1} minimization,'' \emph{Proceedings of
  the National Academy of Sciences}, vol. 100, no.~5, pp. 2197--2202, 2003.
  [Online]. Available: \url{http://www.pnas.org/content/100/5/2197.abstract}
\BIBentrySTDinterwordspacing

\bibitem{CANDES05}
E.~J. Candes and T.~Tao, ``Decoding by linear programming,'' \emph{IEEE
  Transactions on Information Theory}, vol.~51, no.~12, pp. 4203--4215, Dec
  2005.

\bibitem{SELVA17}
J.~Selva, ``M{L} estimation and detection of multiple frequencies through
  periodogram estimate refinement,'' \emph{IEEE Signal Processing Letters},
  vol.~24, no.~3, pp. 249--253, March 2017.

\bibitem{SWINDLEHURST93}
A.~Swindlehurst and T.~Kailath, ``A performance analysis of subspace-based
  methods in the presence of model errors: Part {II} - {M}ultidimensional
  algorithms,'' \emph{IEEE Transactions on Signal Processing}, vol.~41, no.~9,
  pp. 2882--2890, Sept. 1993.

\bibitem{SWINDLEHURST92}
------, ``A performance analysis of subspace-based methods in the presence of
  model errors. {P}art {I}: The {MUSIC} algorithm,'' \emph{IEEE Transactions on
  Signal Processing}, vol.~40, no.~7, pp. 1758--1774, July 1992.

\bibitem{HUBER81}
P.~Huber, \emph{Robust Statistics}.\hskip 1em plus 0.5em minus 0.4em\relax New
  York: John Wiley, 1981.

\bibitem{JACOVITTI94}
G.~Jacovitti and G.~Scarano, ``Hybrid nonlinear moments in array processing and
  spectrum analysis,'' \emph{IEEE Transactions on Signal Processing}, vol.~42,
  no.~7, pp. 1708 -- 1718, July 1994.

\bibitem{STOICA96}
P.~Stoica, M.~Viberg, K.~M. Wong, and Q.~Wu, ``Maximum-likelihood bearing
  estimation with partly calibrated arrays in spatially correlated noise
  fields,'' \emph{IEEE Transactions on Signal Processing}, vol.~44, no.~4, pp.
  888--899, Apr 1996.

\bibitem{GELLI03}
G.~Gelli and L.~Izzo, ``Cyclostationarity-based coherent methods for
  wideband-signal source location,'' \emph{IEEE Transactions on Signal
  Processing}, vol.~51, no.~10, pp. 2471--2482, Oct 2003.

\bibitem{DORON94A}
M.~A. Doron and E.~Doron, ``Wavefield modeling and array processing, part i -
  spatial sampling,'' \emph{IEEE Transactions on Signal Processing}, vol.~42,
  no.~10, pp. 2549--2559, Oct. 1994.

\bibitem{JACOVITTI93}
G.~Jacovitti and G.~Scarano, ``Discrete time techniques for time delay
  estimation,'' \emph{IEEE Transactions on Signal Processing}, vol.~41, no.~2,
  pp. 525--533, Feb. 1993.

\bibitem{MUNKRES57}
J.~Munkres, ``Algorithms for the assignment and transportation problems,''
  \emph{Journal of the Society for Industrial and Applied Mathematics}, vol.~5,
  no.~1, p. 32–38, March 1957.

\bibitem{GRIFFITHS82}
L.~Griffiths and C.~Jim, ``An alternative approach to linearly constrained
  adaptive beamforming,'' \emph{IEEE Transactions Antennas and Propagation},
  vol.~30, no.~1, pp. 27--34, Jan. 1982.

\bibitem{BADDELEY07}
\BIBentryALTinterwordspacing
W.~Weil, Ed., \emph{Spatial Point Processes and their Applications}.\hskip 1em
  plus 0.5em minus 0.4em\relax Berlin, Heidelberg: Springer Berlin Heidelberg,
  2007, pp. 1--75. [Online]. Available:
  \url{https://doi.org/10.1007/978-3-540-38175-4_1}
\BIBentrySTDinterwordspacing

\end{thebibliography}

%\bibliography{SAF}
%\begin{thebibliography}{1}
%\end{thebibliography}

% If you have an EPS/PDF photo (graphicx package needed) extra braces are needed around the contents of the optional argument to biography to prevent the LaTeX parser from getting confused when it sees the complicated \includegraphics command within an optional argument. (You could create your own custom macro containing the \includegraphics command to make things simpler here.)
%\begin{biography}[{\includegraphics[width=1in,height=1.25in,clip,keepaspectratio]{mshell}}]{Michael Shell}

\ifCLASSOPTIONdraftcls
\newpage
\else{\ifCLASSOPTIONtwoside{

%\begin{IEEEbiography}[{\includegraphics[width=1in,height=1.25in,clip,keepaspectratio]{parisi.jpg}}]{Raffaele Parisi}
%received the ``Laurea'' degree in electrical engineering with honors and the Ph.D. degree in information and communication engineering from the University of Rome ``La Sapienza'', Rome, Italy, in 1991 and 1995, respectively.\\
%In 1994 and 1999, he was a Visiting Student and a Visiting Researcher with the Department of Electrical and Computer Engineering, University of California at San Diego, La Jolla. Since 1996, he has been with the University of Rome ``La Sapienza'', where he is currently an Associate Professor with the Department of Information Engineering, Electronics and Telecommunications. His research interests are in the areas of neural networks, digital signal processing, optimization theory, and array processing.
%\end{IEEEbiography}

% You can push biographies down or up by placing a \vfill before or after them. The appropriate use of \vfill depends on what kind of text is on the last page and whether or not the columns are being equalized.

\vfill

% Can be used to pull up biographies so that the bottom of the last one is flush with the other column.
}
\else
\newpage
\fi}
\fi

%%%%%%%%%%%%%%%%%%%%%%%%%%%%%%%%%%%%%%%%%%%%%%%%%%%%%%%%%%%%%%%%%%%%%%%%%%%%%%%%%%%%%%%%%%%%%%%%%%%%%%%%%%%%%%%%%%%%%%%%%%%%%%%%%%%%%%%%%%%%%%%%%%%%%%%%%%%%%%%%%%%%%%%%%%%%%%%%%%%%%%%%%%%%%%%%%%%%

\end{document}